\documentclass{iopart}
\usepackage{iopams}
\usepackage{amssymb}
\usepackage{amsfonts}
\usepackage{graphicx}
\usepackage{bbold}
\usepackage{color}
\usepackage{url}

\eqnobysec

\usepackage{ulem}

\newcommand{\text}[1]{\mbox{#1}}
\newcommand{\leb}[1]{\text{d}[#1]}
\newcommand{\dd}[1]{\text{d}#1}
\newcommand{\Det}{\text{det}}
\newcommand{\Pf}{\text{pf}}
\newcommand{\Sdet}{\text{sdet}}
\newcommand{\Str}{\text{str}}
\newcommand{\Id}{\mathbb{1}}
\newcommand{\IR}{\mathbb{R}}
\newcommand{\IC}{\mathbb{C}}

\newcommand{\IK}{\mathbb{K}}

\newcommand{\nn}{\nonumber}

\newcommand{\diag}{\text{diag}}

\newcommand{\dagg}{^{\dagger}}

\newenvironment{correction}{ \begin{color}{green}}{\end{color}}
\newenvironment{hasto}{\begin{color}{blue}}{\end{color}}

\begin{document}

\title[Distribution of the Smallest Eigenvalue]{Distribution of the Smallest Eigenvalue in Complex and Real Correlated Wishart Ensembles}

\author{Tim Wirtz and Thomas Guhr}

\address{Fakult\"at f\"ur Physik, Universit\"at Duisburg-Essen, Duisburg, Germany}

\ead{\mailto{tim.wirtz@uni-due.de}, \mailto{thomas.guhr@uni-due.de}}

\begin{abstract}
For the correlated Gaussian Wishart ensemble we compute the distribution of the smallest eigenvalue and a related gap probability.We obtain exact results for the complex ($\beta=2$) and for the real case ($\beta=1$). For a particular set of empirical correlation matrices we find universality in the spectral density, for both real and complex ensembles and all kinds of rectangularity. We calculate the asymptotic and universal results for the gap probability and the distribution of the smallest eigenvalue. We use the Supersymmetry method, in particular the generalized Hubbard-Stratonovich transformation and superbosonization.\\\\
\noindent{\it Random Matrix Theory, Supersymmetry, Multivariate Statistics, Correlated Wishart Matrices, Universality, Matrix Model Duality}
\end{abstract}

\pacs{05.45.Tp, 02.50.-r, 02.20.-a}
\ams{62H05}

%\submitto{\JPA}

\maketitle
\section{Introduction}
In modern analysis of complex systems such as communication and information networks, mesoscopic physics, geophysics, biology, financial markets, etc. random matrices play a prominent role \cite{muirhead,Johnstone,ForresterHughes,SanthanamPatra,Seba,TulinoVerdu,Muelleretal,AbeSuzuki,VinayakPandey,LalouxCizeauBouchaudPotters,ple02}, originally introduced by Wishart \cite{Wishart} in the context of biostatistics. He studied ensembles of rectangular random matrices with correlated Gaussian distributed real $\beta=1$ or complex $\beta=2$ entries. Later on  Wigner realized that the spectral fluctuations of a Hamilton operator in the theory of large nuclei can be modeled by Hermitian matrices drawn from a Gaussian distribution providing the same global symmetries  \cite{MehtasBook}. In Ref. \cite{dyson} Dyson showed that there exist three classes of Hermitian random matrices, the Gaussian orthogonal $\beta=1$, the Gaussian unitary $\beta=2$ and the Gaussian symplectic $\beta=4$ ensemble. In random matrix models for Hamiltonian systems, one aims at describing universal spectral fluctuations on the local scale of the mean level spacing. As there is no such scale in most of applications of Wishart random matrices, there is no corresponding universality either. An exception is ``Chiral Random Matrix Theory'', which has much in common with the Wishart random matrix model, it is used to study local universal fluctuations of the Dirac operator \cite{VerbaarschotWettig}.

In the last decades the  connection between multivariate statistics and random matrix theory attracted considerable attention  \cite{muirhead}. Several methods in the classical theory of multivariate statistics such as multivariate analysis of variances, discriminant analysis and principle component analysis are based on the statistical properties, of empirical correlation matrices  \cite{muirhead}. In good agreement with empirical studies of complex systems \cite{Kanasewich,BarnettLewis,Gnanadesikan,chatfield,Johnstone,AbeSuzuki,Seba,LalouxCizeauBouchaudPotters,ple99} it turned out that Gaussian distributed, correlated Wishart matrices provide a realistic and powerful model. As always, the complex case $\beta=2$ is mathematically much easier to treat than the real one $\beta=1$. Thus, although most of the problems in multivariate statistics involve real time series and correlation matrices, exact results are rare. Asymptotic results have the drawback of being given as infinite series in zonal or Jack polynomials, for which resummation in most cases is an unsurmountable task. The difficulty is encoded in an integral over the orthogonal group. It occurs for correlated real Wishart ensembles, reflecting the non invariance of the probability distribution. The integral is known as the ``orthogonal Itzykson-Zuber integral'', or, in mathematics literature as the orthogonal Gelfand spherical function. We show that we can circumvent these difficulties if we employ mutual dualities between matrix models of different dimensions. In some cases, they relate ordinary and ordinary and in other cases ordinary and supermatrix models. 

Owing to the major role of correlation matrices in the analysis of complex systems, it is of no surprise that the extreme eigenvalues are used to study qualitative and quantitative aspects.  The smallest eigenvalue of the Wishart matrix is of considerable interest for statistical analysis, from a general viewpoint and in many concrete applications. In linear discriminant analysis it gives the leading contribution for the \textit{threshold estimate} \cite{LarryWasserman}. It is most sensitive to \textit{noise} in the data \cite{Gnanadesikan}.  In linear principal component analysis, the smallest eigenvalue determines the \textit{plane of closest fit} \cite{Gnanadesikan}.  It is also crucial for the identification of \textit{single statistical outliers} \cite{BarnettLewis}. In numerical studies involving large random matrices, the  \textit{condition number} is used, which depends on the smallest eigenvalue \cite{Edelmann1988,Edelman1992}.  In wireless communication the Multi--Input--Multi--Output (MIMO) channel matrix of an antenna system is modeled by a random matrix \cite{FoschiniGans}. The smallest eigenvalue of $C$ yields an estimate for the \textit{error of a received signal} \cite{UpamanyuMadhow,Burel02statisticalanalysis,ChenTseKahnValenzuela}. In finance, the \textit{optimal portfolio} is associated with the eigenvector to the smallest eigenvalue of the covariance matrix, which is directly related to the correlation matrix \cite{Markowitz}. This incomplete list of examples shows the influence of the smallest eigenvalue in applications. Further information on the role of the smallest eigenvalue is given in \ref{appofsmallestEigenvalue}. It is therefore not only of considerable theoretical interest, but also of high practical relevance  to study its statistics. Our main results are summarized in Ref.~\cite{WirtzGuhrI} . Here we give a detailed derivation addressing also  mathematicians and statisticians as well as further results. 

We exactly calculate the gap probability to find no eigenvalue of a correlated Gaussian distributed Wishart matrix and the distribution of its smallest eigenvalue.  For the real case we find the first time, explicit and easy--to--use formulas for applications. These exact expressions are possible, because of the above mentioned matrix model dualities. There are many studies addressing these issues. For uncorrelated Wishart ensembles exact and asymptotic expression are studied in Refs.~\cite{Edelman1991,GuhrWettigWilke,DamgaardNishigaki,raey,KatzavPerez}. A general discussion of the smallest eigenvalue for arbitrary $\beta$-ensembles is given in Ref.~\cite{RamirezRider}, where the authors showed that it converges in law to the smallest eigenvalue of a stochastic operator. The distribution of the smallest eigenvalue for the complex correlated Wishart ensemble was studied the first time in Ref. \cite{ForresterminEig} and later in Ref. \cite{FangfangZhangYangYang,ZhangNuiYangZahngYang}. In the sequel it was calculated exactly in Ref. \cite{koev}, for all three $\beta$ ensembles. Besides other our results are much easier to handle. Furthermore, we  obtain yet unknown determinant and Pfaffian structures, which amount to a resummation of the results in Ref.~\cite{koev} for the distribution of the smallest eigenvalue. Moreover, we obtain new universalities and of the distribution of the smallest eigenvalue.

We mention that other approaches to study correlations in the Wishart model exist in the literature. To achieve correlations the trace of the Wishart matrix is fixed or an additional average over the variance is introduced. %The former are known as fixed-trace and the latter as norm-dependent ensembles. 
For these ensembles the distribution of the smallest eigenvalue was considered in Refs.~\cite{AkemannVivo2008,ChenDangDa,AkemannVivo2011}. The major difference to our model is that these ensembles have more symmetry and can be studied using orthogonal polynomials.

The article is organized as follows. In section \ref{problem} we give a short sketch of the problem and introduce our notation. Section \ref{dualModels} is concerned with a four-fold duality between different matrix and supermatrix models. These allow us to find exact results for the gap probability in section \ref{exactResults}. In section \ref{asymptotics} we study regimes with  universal spectral fluctuations and the microscopic limit of the gap probability. Both the exact, the asymptotic and universal results for the distribution of the smallest eigenvalue are calculated in section \ref{Mindistribution}, before we compare the analytic results for gap probability and the distribution of the smallest eigenvalue in section \ref{numerics} with numerical simulations. In section  \ref{summary} we summarize  the analytic and asymptotic results and conclude with a list of open problems. 
\section{Formulation of the Problem}\label{problem}
In section \ref{defWE} we define  correlation matrices and discuss how their statistical fluctuations are model using Wishart random matrices. We introduce in section \ref{defEandpmin} the gap probability and discuss its relation to the distribution of the smallest eigenvalue. Section \ref{microscopiclimit} addresses the  microscopic limit.
\subsection{Wishart Model for Correlation Matrices}\label{defWE}
The main area where correlated Wishart random matrices are applied is multivariate statistic  \cite{muirhead,Johnstone}. Suppose we have a set of $p$ time series, all with exactly $n$ time steps, \textit{i.e.}, $X_i\in \IK^n$ for $i=1,\dots,p \leq n$ and $\IK = \IR$ or $\IC$. The entries are denoted by $X_i(t)$, $t=1,\dots,n$. The values of this time series are either real or complex depending on the measured quantity. For a time series $X_i$ with $n$ time steps  we define the sample average to be
\begin{equation}
\left<X_i\right> = \frac{1}{n}\sum_{t=1}^n X_i(t)~.\label{samplaverage}
\end{equation}
To measure the correlations between different time series, one defines the normalized time series
\begin{equation}
  M_i(t) = \frac{X_i(t)- \left<X_i\right>}{\sqrt{\left<X_i^2\right> - \left<X_i\right>^2}}~.
\end{equation}
The Pearson correlation coefficient between the two time series $X_k$ and $X_l$ is given by 
\begin{eqnarray}
\label{entriesC}
  C_{kl} &= \left<M_kM_l^*\right>~,
\end{eqnarray}
where $M_l^*$ is the complex conjugated of the time series $M_l$, for $\beta=2$. If we order the time series $M_i$ into the $p\times n$ dimensional data matrix
\begin{equation}
  M =  \left[\begin{array}{ccc}   M_1(1) & \dots & M_1(n) \\ \vdots &\ddots &\vdots \\ M_p(1) & \dots & M_p(n)
    \end{array}
  \right]~,
\end{equation}
the $p\times p$ sample (or empirical) correlation matrix $C$ is given by
\begin{equation}
  C = \frac{1}{n} MM\dagg~,\label{definitionCmatrix}
\end{equation}
with the entries (\ref{entriesC}). Owing to definition (\ref{definitionCmatrix}), $C$ is positive definite and either real symmetric, if $M$ is real, or  Hermitian, if $M$ is complex. Furthermore, empirical studies showed that the statistical properties of $C$ are quite general consistent with a Gaussian distribution of its entries~\cite{TulinoVerdu,AbeSuzuki,Muelleretal,Seba,SanthanamPatra,LalouxCizeauBouchaudPotters}. Thus, let $W$ be either a real ($\beta=1$) or a complex ($\beta=2$) $p\times n$ matrix, where $n\geq p$. We construct an ensemble of Wishart correlation matrices $WW\dagg/n$ which fluctuates around the empirical correlation matrix $C$. This means that each column vector of the Wishart matrix $W$ follows a multivariate normal distribution with zero mean and correlation matrix $C$. Thus, the probability distribution is \cite{muirhead}
\begin{eqnarray}
 P(W|C) = \frac{1}{(2\pi/\beta)^{pn\beta/2}\Det^{n\beta/2}~C } \exp\left(-\frac{\beta}{2}\tr~WW\dagg C^{-1}\right)
,~\label{WRMdistribution}
\end{eqnarray}
where $\beta=1,2$ denotes the real, respectively the complex Wishart ensemble. $W^\dagger$ is either the transposed of $W$ if $\beta=1$ or the Hermitian conjugate if $\beta=2$. The distribution is normalized,
\begin{equation}
1=\int \leb{W} P(W|C) ~,
\end{equation}
where $\leb{\cdot}$ denotes the flat measure, \textit{i.e}., the product of the independent differentials. By construction we have 
\begin{eqnarray}
C=  \int \leb{W} \frac{1}{n}WW\dagg P(W|C) ~,
\end{eqnarray}
the Wishart correlation matrices $ WW\dagg/n$ yield upon averaging the empirical correlation matrix. From the invariance of the measure $\leb{W}$ it follows that invariant functions,   functions providing an invariance under base changes of its arguments, averaged over the Wishart ensemble, depend on the positive definite ``empirical'' eigenvalues $\Lambda_i$, $i=1,\dots,p$, of $C$ only. We order them in the diagonal matrix $\Lambda$. Thus, if we discuss the joint eigenvalue distribution of $WW\dagg$ in the next section, we can replace $C$ by $\Lambda$.

It is worth mentioning that we are aiming to study the smallest eigenvalue in an ensemble of  correlation matrices fluctuating around a matrix mean value. This is why we refer to $C$ as the empirical correlation matrix, but the results are valid even for $C$ being any positive definite matrix of the symmetry class in question.
\subsection{Distribution of the Smallest Eigenvalue and the Gap Probability}\label{defEandpmin} Let $E^{(\beta)}_{p}(t)$ denote the probability of finding no eigenvalue of $WW\dagg$ within the interval $[0,t]$, referred to as gap probability~\cite{MehtasBook}. In the mathematical literature it is sometimes denoted by $E^{(\beta)}_{p}(0;[0,t])$. It is related to the distribution of the smallest eigenvalue $\mathcal P^{(\beta)}_{\footnotesize\text{min}}(t)$~\cite{MehtasBook} via 
\begin{eqnarray}
 % \fl\quad\quad\quad \quad \quad  
 \mathcal P^{(\beta)}_{\footnotesize\text{min}}(t) = - \frac{\text{d}}{\text{d} t}E^{(\beta)}_{p}(t) \quad \Leftrightarrow \quad E^{(\beta)}_{p}(t) &= 1- \int\limits_0^{t}\dd{t'}\mathcal P^{(\beta)}_{\footnotesize\text{min}}(t') ~. \label{relationpminEgap}
\end{eqnarray}
The gap probability is best expressed in terms of the joint eigenvalue distribution of $WW\dagg$, \textit{i.e.} $P(X|\Lambda)$, where $X=\diag\left(x_1,\dots,x_p\right)$ is the diagonal matrix of eigenvalues of $WW\dagg$. If we diagonalize $WW\dagg= V X V \dagg$ with $V\in \text{U}(p)$ if $\beta=2$ or $V\in \text{O}(p)$ if $\beta=1$, the volume element transforms as
\begin{eqnarray}
 \leb{W} &= \left|\Delta_p(X)\right|^{\beta}\Det^{\beta(n-p+1-2/\beta)/2}X \, \leb{X}\dd{\mu( V)} \ ,
\label{volumeelement}
\end{eqnarray}
where $\dd{\mu}(V)$ is the Haar measure and $\Delta_p(X)=\prod_{i<j}(x_i-x_j)$ is the Vandermonde determinant of  $X$ \cite{MehtasBook}. We introduce
\begin{eqnarray}
\gamma &=& \frac{\beta}{2}(n-p+1) -1 
                = \left\{\begin{array}{ll} (n -p -1 )/2, & \ \beta=1 \\ 
                                                                n-p, & \ \beta=2
                             \end{array}\right. ,
\label{gamma}
\end{eqnarray}
which involves the ``rectangularity'' $n-p$ of the matrix $W$. Substituting this into the Gaussian distribution (\ref{WRMdistribution}) and integrating over either the orthogonal group $O(p)$ if $\beta=1$ or the unitary group if $\beta=2$ leads to the joint distribution of the eigenvalues 
\begin{eqnarray}
P_\beta(X|\Lambda) &=K_{p\times n} \left|\Delta_p(X)\right|^{\beta} \Det^{\gamma}X 
              \, \Phi_\beta(X,\Lambda^{-1}) \ ,
\label{jpdfX}
\end{eqnarray}
with the normalization constant $K_{p\times n}$. We stress that for $\beta=1$ and even rectangularities $n-p$, $\gamma$ is half-integer. Since this leads to certain problems it requires special care. 

The highly non--trivial part in the joint distribution of the eigenvalue (\ref{jpdfX}) is the group integral
\begin{eqnarray}
  \Phi_\beta(X,\Lambda^{-1}) &= \int\dd{\mu( V)}\exp\left(-\frac{\beta}{2} 
   \tr V X V\dagg \Lambda^{-1}\right) \ .
\label{diagdistribution}
 \end{eqnarray}
It is the unitary ($\beta=2$) or the orthogonal ($\beta=1$) Itzykson-Zuber integral. We do not need it explicitly,  we only mention that it is known in the unitary case \cite{HarishChandra,ItzyksonZuber}. Only in the special case of the real spiked Wishart model the orthogonal Izykson-Zuber integral is known due to degeneracies. Then explicit results have been given in Ref.~\cite{MYMo}. Although we do not know the joint eigenvalue distribution in its explicit form, the probability of finding no eigenvalue in an interval of length $t$ including the origin can be written with Eq.~(\ref{diagdistribution}) as \cite{MehtasBook} 
\begin{eqnarray}
E^{(\beta)}_{p}(t) &=  \int\limits\leb{X} P_\beta(X+t\Id_p|\Lambda)~,\label{definitionE}
\end{eqnarray}
where $\Id_p$ is the $p\times p$ dimensional unit matrix and the integration domain is the set of positive diagonal matrices. Formula (\ref{definitionE}) is found by integrating Eq.~(\ref{diagdistribution}) over $[t,\infty)$ and then shifting the eigenvalues $x_i$ by $t$. 
\subsection{Microscopic Limit}\label{microscopiclimit}
Chiral Random Matrix theory, put forward in the context of Quantum Chromodynamics (QCD) in Ref. \cite{Shuryak1993306,VerbaarschotWettig}, is related to the correlated Wishart ensemble with distribution~(\ref{WRMdistribution}). It emerges as a special case of it for $\Lambda = \Id_p$, \textit{i.e.} the entries of Wishart matrix are uncorrelated. It was shown that the quantum fluctuations of the Dirac operator are universal on the scale of the mean level density \cite{BanksCasher,Shuryak1993306,VerbaarschotWettig}. This limit, known as the ``microscopic limit'', is performed by simultaneously scaling the eigenvalues by the mean level spacing and performing the limit $p\rightarrow\infty$. As the local mean level spacing in this regime scales with $1/p$, the microscopic limit is a variant of the unfolding procedure, which is needed to separate system dependent from universal fluctuation on the local scale  \cite{MehtasBook,Guhr1998189}.  In contrast to the microscopic limit in the QCD, we have  to account for the behavior of the empirical eigenvalues $\Lambda$. Universalities of the eigenvalue density were discussed in Ref. \cite{VinayakPandey}. The authors found a criterion to analyze if the level density is universal on the scale of mean level spacing, it is a necessary condition only. We discuss universal regimes  of the correlated Wishart ensemble in section \ref{analysismicroscopiclimit}. There we show that it is meaningful  if we use
\begin{equation}
  t = \frac{u}{4p\eta}\label{scale}
\end{equation}
as ``local scale'', where $\eta$ has to be fixed later on. Hence, we define
\begin{eqnarray}
  \mathcal E^{(\beta)}(u) &= \lim_{p\rightarrow\infty} E^{(\beta)}_{p}\left(\frac{u}{4p\eta}\right)
\end{eqnarray}
and
\begin{eqnarray}
\mathcal  \wp_{\footnotesize\text{min}}^{(\beta)}(u) &= \lim_{p\rightarrow \infty} \frac{1}{4p\eta} \mathcal P_{\footnotesize\text{min}}^{(\beta)}\left(\frac{u}{4p\eta}\right),
\end{eqnarray}
to be the microscopical limit of the gap probability and the distribution of the smallest eigenvalue. Both quantities were already computed for the complex uncorrelated Wishart ensemble (\textit{i.e.} $\Lambda=\Id_p$) in the context of QCD in Ref. \cite{GuhrWettigWilke,DamgaardNishigaki} and in Ref. \cite{KatzavPerez}. For the calculation of the microscopic limit we assume that almost all eigenvalues of $C$ are of the order $O(1)$ and only a finite number are of order $O(p^\tau)$ with $\tau> 0$. This leads to a universality in the spectral fluctuation on the scale of mean level spacing.  Otherwise, in the main part of the study, the eigenvalues are arbitrary. As far as we know, there are no considerations of this kind of microscopic limit for the real correlated Wishart ensembles in the literature.
\section{Mutual Dualities of Matrix Models}\label{dualModels}
Considering the gap probability (\ref{definitionE}), we show that it can be expressed using four different, but mutually dual matrix models in ordinary and superspace. The two in ordinary space are derived in section \ref{fourfoldmatrixduality} and the corresponding dual supermatrix models are constructed in section \ref{susymethod}. Section \ref{synanpsis} summarizes the results schematically.
\subsection{Ordinary Space}\label{fourfoldmatrixduality}
To construct our dual matrix model for the computation of $E^{(\beta)}_p(t)$ we begin with replacing the eigenvalue integral (\ref{definitionE}) by an appropriate Wishart model. The integrand of the gap probability  (\ref{definitionE}) is of the form 
\begin{eqnarray}
\fl\quad \exp\left(-\tr \frac{\beta t}{2\Lambda}\right)  \left|\Delta_p(X)\right|^{\beta}\Det^{\gamma}(X+t\Id_{p})\int\dd{\mu( V)}\exp\left(-\frac{\beta}{2} \tr V X V\dagg \Lambda^{-1}\right)~.\label{integrandE}
\end{eqnarray}
From expression (\ref{integrandE}), there is no unambiguous way to go back to a full matrix model, because there are infinite possibilities to complete the Jacobian in Eq. (\ref{integrandE}). By completing we mean multiplying and simultaneously dividing by a monomial factor in the eigenvalues, 
\begin{equation}
\frac{\prod_{i=1}^p x_i^{\beta(\bar n-p  +1 - 2/\beta)/2}}{\prod_{i=1}^p x_i^{\beta(\bar n-p  +1 - 2/\beta)/2}}= \frac{\prod_{i=1}^p x_i^{\beta(\bar n-p  +1 - 2/\beta)/2}}{\Det^{\beta(\bar n-p  +1 - 2/\beta)/2}X}=1~,\label{monomialfactor}
\end{equation}
to obtain a volume element on the full matrix space of the from Eq.~(\ref{volumeelement}). There are infinitely many possibilities, because the only condition is that the number of columns, corresponding to $\bar n$ in Eq.~(\ref{monomialfactor}), of the full matrix is bigger than $p$. Naively we may insert Eq.~(\ref{monomialfactor}) into integrand~(\ref{integrandE}) with $\bar n =n$ and cast it into the form 
\begin{equation}
\eqalign{
 &  \left|\Delta_p(X)\right|^{\beta}\Det^{\gamma} X~\exp\left(-\tr \frac{\beta t}{2\Lambda}\right)\frac{\Det^{\gamma}(X+t\Id_{p})}{\det^{\gamma} X}\nn\\&\times\int\dd{\mu( V)}\exp\left(-\frac{\beta}{2} \tr V X V\dagg \Lambda^{-1}\right)~,}
\end{equation}
where the number of columns of the underlying full matrix $W$, say, is $n$. Taking the steps of section \ref{defEandpmin} backwards we arrive at the matrix model 
\begin{equation}
\eqalign{
\fl \quad\quad \quad E^{(\beta)}_{p}(t) &=  K_{p\times n}\exp\left(-\tr \frac{\beta t}{2\Lambda}\right) \nn\\\fl \quad\quad \quad &\times \int\leb{W}   \frac{\Det^{\gamma}(WW\dagg+t\Id_{p})}{\det^{\gamma} WW\dagg}\exp\left(-\frac{\beta}{2} \tr WW\dagg\Lambda^{-1}\right)~,} \label{oldE}
\end{equation}
in terms of the $p\times n$ random matrices $W$. We refer to it as the ``large-$W$ model''. The normalization constant $K_{p\times n}$ is chosen properly. This matrix integral is a candidate for applying the Supersymmetry method which  inevitably leads to a supermatrix model. 

Here, we put forward a different approach which will eventually lead us to a much more convenient matrix model in ordinary space. Anticommuting variables will only be used in intermediate steps. The key difference to the approach discussed previously  and the one we propose now is, instead of inserting a factor of one we look for another underlying Wishart matrix model. The Jacobian of the coordinate change $\overline{W}~\overline{W}\dagg \rightarrow VXV\dagg$ should be of the form  
\begin{equation}
\left| \Delta_p(X)\right|^\beta 
\end{equation}
without a monomial factor. Here, $\overline W$ is either a real ($\beta=1$) or  a complex ($\beta=2$) $p\times \bar n$-dimensional Wishart matrix. The number of columns $\bar n\geq p $, of $\overline W$ is a free parameter. It is fixed by the condition that the monomial factor in the corresponding volume element (\ref{volumeelement}), \textit{i.e.} 
\begin{eqnarray}
\Det^{\beta(\bar n-p +1-2/\beta)/2}~X~,
\end{eqnarray}
is unity. From the exponent of the determinant we find
\begin{equation}
\bar n=p-1+\frac{2}{\beta}=p+2-\beta~,\label{definitionNbar}
\end{equation}
for $\beta=1,2$ only. We arrive at the ``small-$W$ model'' dual to the eigenvalue representation of the gap probability in Eq. (\ref{definitionE}),
\begin{equation}
\eqalign{
\fl \quad\quad \quad E^{(\beta)}_{p}(t) &= K_{p\times \bar n} \exp\left(-\tr \frac{\beta t}{2\Lambda}\right) \nn\\\fl \quad\quad \quad &\times \int\leb{\overline W} \Det^{\gamma}(\overline {W}~\overline {W}\dagg+t\Id_{p})\exp\left(-\frac{\beta}{2} \tr\overline W~\overline W\dagg\Lambda^{-1}\right)~,}\label{finalE}
\end{equation}
where we are integrating either over the real or the complex rectangular matrices of dimension $p\times \bar n$. In Eq. (\ref{finalE}) the normalization constant $K_{p\times \bar n} $ is chosen properly. The Wishart model (\ref{finalE}) dual to Eq.~(\ref{definitionE}) has $n-\bar n$ columns less then the naive dual Wishart model (\ref{oldE}). This reduction of number of columns is the crucial difference to the large-$W$ model. In contrast to Eq. (\ref{oldE}), it does not lead to an \textit{averaged ratio} of characteristic polynomials, but just to an \textit{average} of characteristic polynomials. Although this difference is simple, it has dramatic consequences for the dual representation constructed in section \ref{natrualN}.

It is worth mentioning that this is not the only duality. We obtain a more general duality of statistical  quantities in different Wishart matrix models. Let $W$ be a $p\times n$ dimensional Wishart matrix, $l\in \mathbb{N}$ such that  $ n-2 l /\beta \geq p$, $m\in \mathbb{N}$ arbitrary and $f(WW\dagg\Lambda^{-1})$ any smooth, invariant function such that the integral in Eq. (\ref{generalDuality}) exists. Invariant means that $f$ does not change under the transformation $WW\dagg\Lambda^{-1} \rightarrow UWW\dagg\Lambda^{-1}U\dagg$, with either $U\in \text{U}(p)$ if $\beta=2$ or $U\in \text{O}(p)$ if $\beta=1$.  Then we find for an arbitrary $z\in\IC$
\begin{equation}
\eqalign{
\  &\int \leb{W} \frac{\Det^m(WW\dagg + z \Id_{p})}{\Det^l WW\dagg} ~f(WW\dagg\Lambda^{-1})   \\=&\frac{\text{Vol}\left(\text{U}(n)\right)}{\text{Vol}\left(\text{U}(\hat n)\right)}\int \leb{\widehat W} \Det^m(\widehat{W}\widehat{W}\dagg + z \Id_{p}) ~f(\widehat {W}\widehat{W}\dagg\Lambda^{-1})~,}\label{generalDuality}
\end{equation}
where $\widehat{W}$ is either a real ($\beta=1$) or a complex ($\beta=2$) $p\times\hat n$-dimensional matrix, with $\hat n =n-2 l /\beta$.

The dualities we obtained are not the only ones in the literature. In Chiral Random Matrix Theory a duality known as ``flavor-topology duality`` was found \cite{Verbaarschot1994,Verbaarschot,DalmaziVerbaarschot}. It states that the topological charge $\nu=|n-p|$ can be interpreted as $\nu$ additional massless flavor degrees of freedom. For $\beta=2$ this means that
\begin{equation}
 \fl\eqalign{
  Z_{N_f,\nu}^{\beta=2}(m_1,\dots,m_f) &= \int\limits_{\IC^{p\times(p+\nu)}} \leb{W} \prod_{f=1}^{N_f}\Det\left(\mathcal D + m_f^2\Id_{p+n}\right) \exp\left(- \frac{p}{2}\Tr WW\dagg\right)\\ 
  &\sim \int\limits_{\IC^{p\times p}} \leb{W}\Det^{\nu}\mathcal D \prod_{f=1}^{N_f}\Det\left(\mathcal D + m_f^2\Id_{p}\right)  \exp\left(- \frac{p}{2}\Tr WW\dagg\right)\\&= Z_{N_f+\nu,0}^{\beta=2}(m_1,\dots,m_f,0,\dots,0)~,}
  \label{flavortopologyduality}
\end{equation}
where $N_f$ is the number of different flavors, $m_f$ are the flavor masses and
\begin{equation}
 \mathcal D = \left[\begin{array}{cc}0 & \imath W \\\imath W\dagg & 0\end{array}\right]~.
\end{equation}
The relation between the two dualities becomes clear if one compares Eq.~(\ref{flavortopologyduality}) and  Eq.~(\ref{generalDuality}). At the beginning we have a matrix model of dimension $p\times n$ which we reduced to a matrix model of dimension $p\times(n-p-2l/\beta)$. The difference between both dualities is that we start with a determinant in the denominator and therefore decrease the dimensionality. Another crucial difference in our case is the presence of correlations. Thus, our and the flavor-topology duality are based on the same freedom in the Wishart model only.

\subsection{Dual Models in Ordinary and Superspace}\label{susymethod}
Only recently, a matrix model duality was exploited for the correlated Wishart model in Refs.~\cite{RecheretalPRL,Recheretal}. While this dual model is in superspace, we here construct dual models in ordinary and superspace, depending on what turns out to be more convenient. We start with a dual supermatrix model for the case $\gamma\in \mathbb{N}$ and both values of $\beta$ in section \ref{natrualN}.It is derived using two different methods, generalized Hubbard-Stratonovich and Superbosonization. For $\beta=1$ and $n-p$ even we have $\gamma \in \mathbb{N}/2$, this case is treated in section \ref{NoNaturalPower} separately.
\subsubsection{Integer $\gamma$ -- Ordinary Space}\label{natrualN}
The first approach is known as generalized Hubbard-Stratonovich transformation put forward for invariant Hermitian random matrix ensembles in Ref.  \cite{Guhr2006I,KieburgGroenqvistGuhr}. The second approach is  superbosonization and was developed in Ref. \cite{LittlemanSommersZirnbauer}. In Ref. \cite{KieburgSommersGuhr}  the equivalence of both approaches was shown. Both are used, because they have their advantages and disadvantages if various types of limits are considered. We start with the generalized Hubbard-Stratonovich transformation. We do not present the details of the approaches, because there are many reviews in the literature \cite{Guhr2006I,KieburgGroenqvistGuhr,LittlemanSommersZirnbauer,GuhrSUSY,SUSYZirnbauer}. For an application to correlated Wishart models we refer to \cite{RecheretalPRL,Recheretal}. 
\paragraph{Generalized Hubbard-Stratonovich Transformation}
The matrix model considered here is the small-$W$ model of Eq.~(\ref{finalE}). For the case of an integer power $\gamma$ in Eq.~(\ref{finalE}), we derive a matrix model in ordinary space. The random matrix belongs either to the  unitary or the symplectic ensemble. If  $\beta=1$ it is a $\gamma\times\gamma$ self-dual Hermitian matrix with quaternion entries and if $\beta=2$ it is Hermitian.

For the convenience of the reader, we sketch the salient features of the generalized Hubbard-Stratonovich transformation applied to the present case in \ref{generalizedHS}. We obtain the following expression for the gap probability,
 \begin{equation}
\eqalign{
  E^{(\beta)}_{p}(t)&=K_{p\times\bar n} \exp\left(-\tr\frac{\beta t}{2\Lambda}\right) \int\leb{\sigma} \exp\left(-\tr \sigma \right)f_{\bar n ,\beta}(\sigma) \\ 
&\times\prod_{k=0}^p\Det^{\beta/2}\left(\frac{\beta t}{2}\Id_{ 2\gamma/\beta}-\Lambda_k\sigma\right)~,}\label{finalESUSY}
\end{equation}
where
\begin{eqnarray}
f_{\bar n,\beta}(\sigma) &= \int\leb{\varrho}\Det^{\beta \bar n /2}\varrho~\exp\left(-\imath \tr\varrho\sigma\right) ~.\label{backtransIngSieg}
\end{eqnarray}
The domain of integration in $f_{\bar n,\beta}(\sigma)$ and (\ref{finalESUSY}) are the same such that $\varrho$ is $2\gamma/\beta\times2\gamma/\beta$-dimensional and either a real quaternion self-dual or a Hermitian matrix. Hence, there are no Grassmanian variables neither in Eq.~(\ref{finalESUSY}) nor in  Eq.~(\ref{backtransIngSieg}).

Since $\sigma$ is either a Hermitian ($\beta=2$) or a self-dual Hermitian matrix with quaternion entries ($\beta=1$) it is diagonalizable, \textit{i.e.} $\sigma=usu\dagg$, where $u\in\text{USp}(2\gamma)$ if $\beta=1$ or $u \in \text{U}(\gamma)$ if $\beta=2$ and $s=\Id_{2/\beta}\otimes\diag(s_1,\dots,s_\gamma)$. Due to the invariance of $f_{\bar n,\beta}(\sigma)$ and the integration measure $\leb{\sigma}$ under the action of $\text{USp}(2\gamma)$ or $\text{U}(\gamma)$ in Eq. (\ref{finalESUSY}), we have to deal with $\gamma$ eigenvalue integrals only.
\paragraph{Superbosonization}
The main difference compared to the generalized Hubbard-Stratonovich transformation is the integration domain. But both results can be transformed into each other \cite{KieburgSommersGuhr}. Transferring the steps taken in Ref.~\cite{Recheretal} to our case we find for the gap probability
\begin{equation}
\eqalign{
 E^{(\beta)}_{p}(t) &=K_{p\times\bar n}\exp\left(-\tr\frac{\beta t}{2\Lambda}\right) t^{\gamma(\beta-2)} \int\leb{U}\Det^{-\kappa}U\\&\times\exp\left(-\imath\frac{\beta t }{2}\tr U\right)\prod_{k=1}^{p} \Det^{ -\frac{\beta}{2}}\left(\Id_{2\gamma/\beta}-\imath\Lambda_k U\right)~,}\label{SuperbosonizationE}
\end{equation}
where we defined $\kappa = \beta \bar n/2+\gamma +\frac{\beta-2}{2}=\beta p/2+\gamma$. The domain of integration is either given by $\text{U}(\gamma)$ or by $\text{USp}(2\gamma)$ for $\beta=2$ respectively $\beta=1$. We mention that the measure $\leb{U}$ is the usual flat one, but the Haar measure is obtained if one combines the determinant of $U$ to a power of $-\gamma -\frac{\beta-2}{2}$ with $\leb{U}$,
\begin{equation}
\dd{\mu(U)} \sim \leb{U}\Det^{-\gamma -\frac{\beta-2}{2}}U~. 
\end{equation}
Thus we are in the situation of the circular unitary ensemble  for $\beta=2$ and circular symplectic ensemble  for $\beta=1$. As mentioned above, the normalization constant is yet to be determined and has to be distinguished from the one of generalized Hubbard-Stratonovich transformation.
\subsubsection{Half-Integer $\gamma$ -- Superspace}\label{NoNaturalPower}
Up to now we restricted our analysis to an integer power $\gamma$ of the determinant in Eq. (\ref{finalE}). But for $\beta=1$, $\gamma = (n-p+1)/2 -1$ can be half-integer. We extend our analysis to half-integer power $\gamma = \alpha +1/2$, where $\alpha=(n-p-2)/2\in\mathbb N$ and we assume that $\alpha>0$. In what follows we stress only the differences between this calculation and the one of section \ref{susymethod}.
For this particular case we use the Hubbard-Stratonovich transformation only. 
\\\\The key to obtain a ratio of determinants that can be handled with generalized Hubbard-Stratonovich transformation, is to extend the integrand in Eq. (\ref{finalE}). We cast the determinant of half-integer power into the form
\begin{eqnarray}
 \Det^{\alpha + 1/2}\left(WW\dagg + t \Id_p\right)&= \frac{\Det^{\alpha + 1}\left(WW\dagg + t \Id_p\right)}{\Det^{1/2}\left(WW\dagg + t \Id_p\right)}~.
\label{GammahalfDet}
\end{eqnarray} 
This ratio of characteristic polynomials can be handled with supersymmetry. For the details we refer to  \ref{generalizedHS}. The supermatrix model we obtain is somewhat unusual. Therefore, it is better to give a particular parametrization and the form of the flat measure.  The supermatrix $\mu$ is given by
\begin{eqnarray}
  \mu &= \left[\begin{array}{cc} y & \begin{array}{cc} \eta\dagg & \eta^T \end{array} \\ \begin{array}{c} \eta \\-\eta^* \end{array} &  \imath\sigma \end{array}\right]~.\label{parametrizationmu}
\end{eqnarray}
where $\eta$ is a $(\alpha+1)$-dimensional complex Grassmannian, $y$ is a real number and $ \sigma$ is a $(2\alpha+2)\times (2\alpha+2)$-dimensional self-dual Hermitian matrix with quaternion entries \cite{Guhr2006I,KieburgGroenqvistGuhr}. A Wick rotation of $\mu$ is performed, ensuring the convergence of the superintegral and leading to $\imath \sigma$ in Eq.~(\ref{parametrizationmu}). The flat measure on the superspace reads
\begin{eqnarray}
  \leb{\mu} = \leb{\sigma}\dd{y}\prod^{\gamma}_{i=1}\dd{\eta_i^*}\dd{\eta_i}~.
\end{eqnarray}
The supermatrix model for the gap probability is then
\begin{equation}
\eqalign{   E^{(1)}_{p}(t) &= K_{p\times\bar n}\exp\left(-\tr\frac{t}{2\Lambda}\right)\int\leb{\mu}\exp\left(-\Str \mu\right)~ I_{\beta}(\mu)\\&\times \prod_{k=1}^p\Sdet^{-1/2}\left(\frac{t}{2}\Id_{2\alpha+3}-\Lambda_k\mu\right)~,}\label{finalhalfpowerexponent}
\end{equation}
with the  supersymmetric Ingham-Siegel integral of Eq. (\ref{Ingham-Siegel}). The integration domain and the parametrization of $\nu$ in  Eq. (\ref{Ingham-Siegel}) are those  of $\sigma$ in Eq. (\ref{finalhalfpowerexponent}).
\subsection{Synopsis} \label{synanpsis}
We obtain altogether four dual matrix models. Two ordinary ones of Wishart type, the large-$W$ and  the small-$W$ model in section \ref{fourfoldmatrixduality} as well as a dual ordinary invariant matrix model, the small-$\sigma$ model in section \ref{susymethod}. But there exists a fourth dual model. The large-$\sigma$ supermatrix model. It is achieved if one applies the machinery of section \ref{susymethod} and \ref{generalizedHS} to the large-$W$ model. This mutual four-fold duality does not hold for integer $\gamma$ only, but also if $\gamma$ is half-integer. We summarize this four-fold duality schematically
 \begin{equation*}
\eqalign{
 \fl\quad\quad\quad   &\nearrow \begin{array}{c} \text{\textbf{large-$W$ model}}\\p\times n\text{-dim.} \end{array} \Leftrightarrow \begin{array}{c} \text{\textbf{large-$\sigma$ model}}\\(2\gamma\beta|2\gamma\beta)\times (2\gamma\beta|2\gamma\beta)\text{-dim.}\end{array}\\
\fl\quad\quad\quad E^{(\beta)}_{p}(t)& \\
\fl\quad\quad\quad &\searrow \begin{array}{c}\text{\textbf{small-$W$ model}}\\p\times\bar n\text{-dim.}\end{array}\Leftrightarrow\quad\quad~\begin{array}{c} \text{\textbf{small-$\sigma$ model}}\\2\gamma/\beta\times 2\gamma/\beta\text{-dim.}\end{array}}\nn
 \end{equation*}
if $\gamma \in \mathbb{N}$ and
 \begin{equation*}
\eqalign{
 \fl\quad\quad\quad   &\nearrow \begin{array}{c} \text{\textbf{large-$W$ model}}\\p\times n\text{-dim.} \end{array} \Leftrightarrow \begin{array}{c} \text{\textbf{large-$\sigma$ model}}\\(2\alpha +2|2\alpha+2)\times (2\alpha+2|2\alpha+2)\text{-dim.}\end{array}\\
\fl\quad\quad\quad E^{(\beta)}_{p}(t)& \\
\fl\quad\quad\quad &\searrow \begin{array}{c}\text{\textbf{small-$W$ model}}\\p\times\bar n\text{-dim.}\end{array}\Leftrightarrow\quad\quad~\begin{array}{c} \text{\textbf{small-$\sigma$ model}}\\(1|2\alpha+2)\times (1|2\alpha+2)\text{-dim.}\end{array}}
 \end{equation*}
if $\beta=1$ and $\gamma \in \frac{1}{2} \mathbb{N}$ with $\gamma = (2\alpha +1)/2$. It should be emphasized that this scheme is true for all kinds of invariant probability distributions. This is a consequence of the arguments leading to the correspondence in Eq. (\ref{generalDuality}) and the generalized Hubbard-Stratonovich transformation of \ref{generalizedHS}.
\section{Exact Results}\label{exactResults}
The small-$\sigma$ matrix model~(\ref{finalESUSY}) obtained in section \ref{natrualN} for integer $\gamma$, depends on the eigenvalues of $\sigma$ only. Thus, we can  diagonalize the integration measure such that we are  left with integrals over the eigenvalues. Since it is an ordinary Hermitian matrix model, no Efetov-Wegner or Rothstein term occur. Due to the distributive nature of $f_{\bar n , \beta}(\sigma)$, the eigenvalue integrals are trivial. For superbosonization we will find contour integrals over the eigenvalues, which can be done using the residue theorem. Since both approaches are equivalent we will see that both lead to the same result. We start in section \ref{exactSol} with the results of the generalized Hubbard-Stratonovich transformation and discuss in section \ref{exactSB} the approach using superbosonization.
\subsection{Models Derived Using the Generalized Hubbard-Stratonovich}\label{exactSol}
We solve the dual small-$\sigma$ (\ref{finalESUSY}) with standard methods. Starting point is the complex case in section  \ref{exactcomplex}. In section \ref{exactreal}  we adapt the calculations to the real case.
\subsubsection{Complex Case}\label{exactcomplex}
Since the integral $f_{\bar n,2}(\sigma)$ is invariant under the action by conjugation of an element of $\text{U}(\gamma)$, we can diagonalize the $\sigma$ integral, \textit{i.e.} $\sigma=usu\dagg$. Here is $u\in \text{U}(\gamma)$ and $s=\text{diag}(s_1,\dots,s_\gamma)$ is the matrix of eigenvalues. The integration domain is the space of ordinary Hermitian matrices. Hence, diagonalization does not lead to a boundary  or Efetov-Wegner term. 

In Ref.~\cite{Guhr2006I} the author showed, by a direct calculation, that $f_{\bar n,2}(\sigma)$ is proportional to  derivatives of a delta function.  We will give a short sketch how to calculate this integral in the complex case, \textit{i.e.}
\begin{eqnarray}
\fl \quad\quad\quad\quad\quad f_{\bar n, 2}(s) &= \int\leb{\varrho}\Det^{\bar n}\varrho~\exp\left(-\imath \tr\varrho s\right)~.
\end{eqnarray}
Diagonalizing the integration measure, \textit{i.e.}, $\varrho=vrv\dagg$  where $v\in \text{U}(\gamma)$, leads to a Jacobian given by the squared  Vandermonde determinant of $r$. The flat measure on the space of Hermitian matrices decomposes into the flat measure on the space of eigenvalues times a Vandermonde determinant and the Haar measure on $\text{U}(\gamma)$. Substituting this into the integral representation of $f_{\bar n,2}(s)$ we find
\begin{equation}
\eqalign{
\fl \quad\quad\quad\quad\quad  f_{\bar n, 2}(s) &= \int\limits_{\IR^{\gamma}}\leb{r}~\Delta^2_{\gamma}(r) \Det^{\bar n}r~\int\limits_{\footnotesize\text{U}(\gamma)}\dd{\mu(v)}~\exp\left(-\imath\tr~ vrv\dagg s\right)}~. \label{InghamSiegelDist}
\end{equation}
The group integral is the Harish-Chandra-Itzykson-Zuber integral. It is known for the unitary group only. An exact solution can be found in Ref. \cite{ItzyksonZuber,HarishChandra}, and is given by
\begin{eqnarray}
\fl\quad\quad\quad\quad\quad\int\limits_{\footnotesize\text{U}(\gamma)}\dd{\mu(v)} ~\exp\left(-\imath\tr v r v\dagg s\right) &\sim\frac{\Det ~\exp\left(-\imath r_i s_j\right)}{\Delta_{\gamma}(r)\Delta_{\gamma}(s)} ~.
\end{eqnarray}   
Expanding $\Det \left[\exp\left(-\imath r_i s_j\right)\right]$ and performing appropriate changes of integration variables leads to 
\begin{eqnarray}
\fl\quad\quad\quad\quad\quad f_{\bar n,2}(s) &\sim \frac{1}{\Delta_{\gamma}(s)}\int\limits_{\IR^{\gamma}}\leb{r}~ \Delta_{\gamma}(r) \Det^{\bar n}r ~\exp\left(-\imath \tr sr\right) ~. 
\end{eqnarray}
The determinant to the power of $\bar n$ and the Vandermonde determinant of $r$ can be expressed as one determinant of derivatives with respect to the components of $s$\begin{eqnarray}
\fl\quad\quad\quad\quad\quad  f_{\bar n,2}(s) &\sim \frac{1}{\Delta_{\gamma}(s)}\Det \left[\frac{\partial^{\bar n+j-1}}{\partial s_i^{\bar n+j-1}} \delta(s_i) \right]\int\limits_{\IR^{\gamma}}\leb{r}~ \exp\left(-\imath \tr sr\right) ~. \label{diffdeltaint}
\end{eqnarray}
Using the properties of the delta function as distribution we can cast Eq. (\ref{diffdeltaint}) into the form
\begin{eqnarray}
\fl\quad\quad\quad\quad\quad  f_{\bar n,2}(s) & \sim \prod_{i=1}^{\gamma}\frac{\partial^{\bar n+\gamma-1}}{\partial s_i^{\bar n+\gamma-1}} \delta(s_i) ~.\label{InghamSiegelDistResult}
\end{eqnarray}
This way of computations works for the complex case only, because the Harish-Chandra-Itzykson-Zuber integral is known. An alternative way of deriving Eq.~(\ref{InghamSiegelDistResult}) from Eq.~(\ref{InghamSiegelDist}) is given in Ref.~\cite{KieburgGroenqvistGuhr}, where the authors use linear differential operators in $s$. If we substitute Eq.~(\ref{InghamSiegelDistResult}) into Eq.~(\ref{finalESUSY}) and diagonalize $\sigma$, Eq.~(\ref{finalESUSY}) reduces to 
\begin{eqnarray}
\fl\quad\quad\quad\quad\quad E^{(2)}_{p}(t)&=K_{p\times \bar n}\exp\left(-\tr\frac{t}{\Lambda}\right) \int\limits_{\IR^{\gamma}}\leb{s}~\Delta^2_{\gamma}(s)~\prod_{j=1}^{\gamma}w(s_j,t)~
\label{beta2NoDet}~,
\end{eqnarray}
where we defined the weight function
\begin{eqnarray}
\fl\quad\quad\quad\quad\quad w(z,t)&=\exp\left(- z\right)~\prod_{k=1}^p\left(t-\Lambda_k z\right) \frac{\partial^{\bar n+\gamma-1}}{\partial z^{\bar n+\gamma-1}} \delta(z)~.
\end{eqnarray}
Combining standard techniques \cite{MehtasBook} and the results of Ref.~\cite{KieburgGuhrI}, we express the gap probability as determinant,
\begin{eqnarray}
\fl\quad\quad\quad\quad E^{(2)}_{p}(t)=K_{p\times \bar n}\exp\left(-\tr\frac{t}{\Lambda}\right)\Det\left[\int\limits_{-\infty}^{\infty}\dd{z}~w(z,t)~z^{j-1+i-1}\right]_{i,j=1,\dots,\gamma}~. \label{beta2determinantE}
\end{eqnarray}
The $p$-fold product in the weight function $w(z,t)$ can be written as polynomial in $z$ with elementary symmetric polynomials $ e_k(\Lambda)$ as coefficients  \cite{KraftProcesi}
\begin{eqnarray}
\prod_{k=0}^p\left(t-\Lambda_k z\right) &= \sum_{k=0}^p (-1)^k t^{p-k} e_k(\Lambda) z^k ~,\label{SeriesExp}
\end{eqnarray}
where $e_k$ denotes the $k$th elementary symmetric function. It reads
\begin{eqnarray}
e_k(\Lambda) &= \sum_{\scriptscriptstyle 1\leq i_1 <\dots< i_k \leq p} \Lambda_{i_1}\cdots\Lambda_{i_k}~.
\end{eqnarray}
For example, the first three elementary symmetric functions are 
\begin{eqnarray}
e_0(\Lambda) &\equiv 1~,\\
e_1(\Lambda) &= \Lambda_1 + \Lambda_2 + \dots + \Lambda_p~,\\
e_2(\Lambda) &= \Lambda_1\Lambda_2 + \Lambda_1\Lambda_3 +\dots+\Lambda_{p-1}\Lambda_p~.
\end{eqnarray}
The $z$-integral is  given by the derivatives of the integrand at zero, \textit{i.e.}
\begin{eqnarray}
\fl\int\limits_{-\infty}^{\infty}\dd{z}~w(z,t)~z^{j-1+i-1} &\sim \sum_{k=0}^p (-1)^k t^{p-k} e_k(\Lambda) \left.\frac{\partial^{\bar n+\gamma-1}}{\partial z^{\bar n+\gamma-1}}\right|_{z\rightarrow 0} z^{k+j+i-2} \exp(- z)\\
\fl\quad\quad&\sim \Theta(\alpha_{p,2})(-1)^{i+1} \sum_{k=0}^{\text{min}(p,\alpha_{p,2})} \frac{ e_k(\Lambda)~  t^{p-k}}{(\alpha_{p,2}-k)!}~.  \label{complexdeltaintegral}
\end{eqnarray}
In the expression above we defined  $\alpha_{p,2} = p+\gamma+1-i-j$ and use the Heaviside function   $\Theta(x)$. Substituting the expression above for the determinant kernel into Eq.~ (\ref{beta2determinantE}) yields
\begin{eqnarray}
\fl\quad  E^{(2)}_{p}(t)=\frac{\exp\left(-\displaystyle\tr\frac{t}{\Lambda}\right)}{\Det^\gamma\Lambda }\Det\left[\Theta(\alpha_{p,2})(-1)^{i+1} \sum_{k=0}^{\text{min}(p,\alpha_{p,2})} \frac{ e_k(\Lambda)~  t^{p-k}}{(\alpha_{p,2}-k)!}\right]_{i,j=1,\dots,\gamma}  \label{finalEbeta2}~,
\end{eqnarray}
where we already insert the correct normalization 
\begin{eqnarray}
K_{p\times \bar n} = \Det^{-\gamma}\Lambda, 
\end{eqnarray}
which was computed using the expression found by inserting Eq.~(\ref{complexdeltaintegral}) into Eq.~(\ref{beta2determinantE}) and the requirement $\lim_{t\rightarrow 0} E^{(2)}_{p}(t)=1$. 
\subsubsection{Real Case}\label{exactreal}
Although the real case is much more involved compared to the complex, it is even here possible to calculated  the gap probability exactly. The main difficulty  is to compute the integral $ f _{\bar n,1}(\sigma)$. In the same manner as in the complex case it will lead to a distribution or rather to the derivatives of  delta functions. It  was calculated in Ref.~\cite{KieburgGroenqvistGuhr}, and can be written as
\begin{eqnarray}
 f_{\bar n,1}(s) &\sim \prod_{i=1}^{\gamma} \frac{\partial^{\bar n+2\gamma -2 }}{\partial s_i^{\bar n+2\gamma-2}}~\delta(s_i)~,\label{InghamSiegelDistResultbeta1}
\end{eqnarray}
where $s_1,\dots,s_\gamma$ are the distinct eigenvalues of $\sigma$ order in the diagonal matrix $s$ and $\bar n= p+1$. The proportionality constant is absorbed into the overall constant of the observable. Substituting Eq.~(\ref{InghamSiegelDistResultbeta1}) into Eq.~(\ref{finalESUSY}) and diagonalizing the self-dual, Hermitian matrix $\sigma$ yields 
\begin{eqnarray}
\fl \quad\quad\quad E^{(1)}_{p}(t)&=K_{p\times \bar n}~ \exp\left(-\tr\frac{t}{2\Lambda}\right) \int\limits_{\IR^{\gamma}}\leb{s}~\Delta_{\gamma}^4(s)~\prod_{i=1}^{\gamma} ~w(s_i,t)~.
\end{eqnarray}
The Vandermonde determinant to the power of four is the Jacobian coming from the diagonalization of $\sigma$. For the shake of compactness we defined the weight function
\begin{eqnarray}
w(z,t) = \prod_{k=1}^p\left(\frac{t}{2}-\Lambda_k z\right)~\exp\left(- 2z\right)\frac{\partial^{\bar n+2\gamma-2 }}{\partial z^{\bar n+2\gamma-2}}\delta(z)~.
\end{eqnarray}
The problem of solving the eigenvalue integral is straightforward. We obtain a Pfaffian compared to the determinant in the complex case,
\begin{eqnarray}
\fl\quad E^{(1)}_{p}(t) &=K_{p\times \bar n}~ \exp\left(-\tr\frac{t}{2\Lambda}\right)   \Pf\left[\int\limits_{-\infty}^\infty\dd{z}~z^{i+j-3}~(j-i)~w(z,t)\right]_{i,j=1,\dots, 2\gamma}.
\end{eqnarray}
Using Eq.~(\ref{SeriesExp}) we find
\begin{eqnarray}
\fl\quad\quad\quad\quad
\int\limits_{-\infty}^\infty\dd{z}~z^{i+j-3}~w(z,t)
&\sim\Theta(\alpha_{p,1})(-1)^{i+j}\sum^{\text{min}(p,\alpha_{p,1})}_{k=0}\frac{e_k(\Lambda)}{(\alpha_{p,1}-k)!}t^{p-k}~,
\end{eqnarray}
where we introduce the constant $\alpha_{p,1}=p  +2\gamma+2-i-j$. As one might have expected, the gap probabilities in the real and the complex case have much in common, in particular the kernels look quite similar. The  full and exact expression for the gap probability is 
\begin{equation}
\eqalign{
\fl\quad\quad\quad\quad E^{(1)}_{p}(t) &= \frac{\exp\left(-\displaystyle\tr\frac{t}{2\Lambda}\right)}{\Det^{\gamma}\Lambda}\nn \\\fl\quad\quad\quad\quad&\times \Pf\left[(j-i)\Theta(\alpha_{p,1})(-1)^{i+j}\sum^{\text{min}(p,\alpha_{p,1})}_{k=0}\frac{e_k(\Lambda)~t^{p-k}}{(\alpha_{p,1}-k)!}\right]_{i,j=1,\dots 2\gamma}.}\label{FinalExactEbeta1}
\end{equation}
The normalization constant
\begin{eqnarray}
K_{p\times \bar n}= \Det^{-\gamma}\Lambda~ .\label{beta1normal}
\end{eqnarray}
was computed utilizing the requirement $\lim_{t\rightarrow 0} E^{(1)}_{p}(t) =1$.
\subsection{Models Derived Using Superbosonization}\label{exactSB}
Although it was shown in Ref. \cite{KieburgSommersGuhr} that superbosonization and generalized Hubbard-Stratonovich are equivalent, we compute the gap probability also with the help of superbosonization. We discuss the complex and the real case in section \ref{superbosonizationexactcomplex} and \ref{superbosonizationexactreal}, respectively.
\subsubsection{Complex Case}\label{superbosonizationexactcomplex}
 Consider Eq.~(\ref{SuperbosonizationE}), it follows from Ref. \cite{LittlemanSommersZirnbauer} that the integration domain is the unitary group $\text{U}(\gamma)$. The CUE of Eq.~(\ref{SuperbosonizationE}) is invariant under the adjoint action of $\text{U}(\gamma)$, such that we can diagonalize it with a Jacobian of the form $\Delta^{2}_{\gamma}(s)$. In  $s$ we order the eigenvalue of the unitary matrix $U$. Hence diagonalization of the integral of Eq.~(\ref{SuperbosonizationE}) yields 
\begin{equation}
\eqalign{
\fl \quad\quad\quad\quad E^{(2)}_{p}(t) &= K_{p\times\bar n} \exp\left(-\tr\frac{ t}{\Lambda}\right)  \prod_{j=1}^{\gamma}~\oint\limits_{|s_j|=1}\dd{s_j}~\Delta^{2}_{\gamma}(s)\Det^{-\kappa}s\nn\\\fl\quad\quad\quad\quad&\times\exp\left(-\imath t \tr s\right) \prod_{k=1}^p \Det\left(\Id_{\gamma} -\imath \Lambda_k s \right)~.}
\end{equation}
Where $\kappa = \bar n + \gamma$.  Due to the scaling invariance of closed contour integrals we can rescale $s$ by  $-\imath $ and obtain
\begin{equation}
\eqalign{
\fl \quad\quad\quad\quad E^{(2)}_{p}(t) &= K_{p\times\bar n} \exp\left(-\tr\frac{t}{\Lambda}\right)  \prod_{j=1}^{\gamma}~\oint\limits_{|s_j|=1}\dd{s_j}~\Delta^{2}_{\gamma}(s)\Det^{-\kappa}s\nn\\\fl\quad\quad\quad\quad&\times\exp\left(-t \tr s\right) \prod_{k=1}^p \Det\left(\Id_{\gamma} - \Lambda_k s \right)~.}
\end{equation}
Standard textbook techniques  \cite{MehtasBook,KieburgGuhrII} can be used to show that $E^{(2)}_{p}(t)$ has a determinantal structure. We find
\begin{equation}
\eqalign{
\fl\quad\quad\quad\quad  E^{(2)}_{p}(t) &= 
K_{p\times\bar n} \exp\left(-\tr\frac{t}{\Lambda}\right) \nn\\\fl\quad\quad\quad\quad &\times \Det\left[\oint\limits_{|z|=1}\dd{z}~z^{i+j-2-(p+\gamma)}~\exp\left(-tz\right) \prod_{k=0}^p \left(1 - \Lambda_kz\right)\right]_{ i,j=1,\dots\gamma}}.\label{startpointasymptoticscomplex}
\end{equation}
The integral in the determinant kernel has a pole for all values of $i,j$, except for instance if $i+j>\gamma+2$ and $p\leq\gamma$. Thus we can use residue theorem to compute it and by doing so  we find that the determinant kernel is given by 
\begin{equation}
\eqalign{
&\oint\limits_{|z|=1}\dd{z}~z^{i+j-p-2-\gamma}~\exp\left(-tz\right) \prod_{k=0}^p \left(1 - \Lambda_kz\right) \nn\\
&\sim\Theta(\alpha_{p,2}) \sum^{\text{min}(p,\alpha_{p,2})}_{k=0}\frac{(-1)^{i+1}e_{k}(\Lambda)~ t^{\alpha_{p,2}-k}}{(\alpha_{p,2}-k)!}~.}
\end{equation}
Substituting this into the expression found earlier for $E^{(2)}_{p}(t)$ yields the gap probability of the $\beta=2$ ensemble. It is of the form 
\begin{eqnarray}
\fl E^{(2)}_{p}(t) &= K_{p\times\bar n} \exp\left(-\tr\frac{t}{\Lambda}\right) \Det\left[\Theta(\alpha_{p,2})(-1)^{i+1}\sum^{\text{min}(p,\alpha_{p,2})}_{k=0}\frac{e_{k}(\Lambda)~t^{\alpha_{p,2}-k} }{(\alpha_{p,2}-k)!}\right]_{i,j=1,\dots,\gamma}~.\label{notnormalizedsuperbos1}
\end{eqnarray}
With the aid of Eq.~(\ref{notnormalizedsuperbos1}) it is  possible to determine the unknown normalization constant. Employing the condition $\lim_{t\rightarrow 0} E^{(2)}_{p}(t) =1$ fixes it. Applying it to the expression above yields
\begin{eqnarray}
\lim_{t\rightarrow 0} E^{(2)}_{p}(t) &=   K_{p\times\bar n} \Det^{\gamma}\Lambda ~.
\end{eqnarray}
Hence, we succeeded in giving an exact formula for the normalized gap probability $\beta=2$,
\begin{eqnarray}
\fl \quad E^{(2)}_{p}(t) &=  \frac{\displaystyle\exp\left(-t\tr\Lambda^{-1}\right)}{\Det^{\gamma}(\Lambda)} \Det\left[\Theta(\alpha_{p,2})(-1)^{i+1} \sum^{\text{min}(p,\alpha_{p,2})}_{k=0}\frac{e_{k}(\Lambda)~t^{\alpha_{p,2}-k} }{(\alpha_{p,2}-k)!} \right]_{ i,j=1,\dots\gamma}. \label{Ebeta2SuperB}
\end{eqnarray}
To see the connection between both approaches there are two possibilities, either reorganizing the determinant or by going back to Eq.~(\ref{finalESUSY}). If we rescale $\sigma$ by $t$ and use that $I_{\beta}(t\sigma)=t^{-\gamma(\bar n+\gamma)}~I_{\beta}(\sigma)$ we obtain, after solving the matrix integral, Eq.~(\ref{Ebeta2SuperB}).
\subsubsection{Real Case}\label{superbosonizationexactreal}
The arguments of invariance under the action of $\text{USp}(2\gamma)$ go through as in the complex case. Diagonalization of Eq.~(\ref{SuperbosonizationE}), with a Jacobian $\Delta_{\gamma}^{4}(s)$, yields 
\begin{equation}
\eqalign{
\fl \quad\quad E^{(1)}_{p}(t) &= K_{p\times\bar n} \exp\left(-\tr\frac{t}{2\Lambda}\right) t^{-\gamma}  \nn\\\fl\quad\quad&\times \prod_{i=1}^\gamma~\oint\limits_{|s_i|=1}\dd{s_i}\Det^{-\kappa}s~\exp\left(-\imath t\tr s\right)\prod_{k=1}^{p} \Det\left(\Id_{2\gamma}-\imath\Lambda_k s\right) }
~,\label{startingexactEbeta2SB}
\end{equation}
where $\kappa=\bar n/2+\gamma-1/2=p/2 + \gamma$. By the same arguments as above it is allowed to rescale the contour integral by $-\imath$. Hence we get rid of the $\imath$ in front of the $s$. As mentioned above we can stop our calculations here if we rescale the eigenvalues by $1/t$. The $t$ dependence of Eq.~(\ref{startingexactEbeta2SB}) is then similar to the model obtained using generalized Hubbard-Stratonovich. It turns out that Eq.~(\ref{startingexactEbeta2SB}) can be brought to the form
\begin{equation}
\eqalign{
\fl\quad\quad E^{(1)}_{p}(t) &= K_{p\times\bar n} \exp\left(-\tr\frac{t}{2\Lambda}\right)\nn\\\fl\quad\quad&\times\Pf\left[\oint\limits_{|z|=1}\dd{z} ~z^{i+j-3-p-2\gamma}~(j-i)~\exp\left(-z\right)\prod_k^{p}(t-\Lambda_kz)\right]_{i,j=1,\dots, 2\gamma}}~.\label{startpointasymptoticsreal}
\end{equation}
Applying the  residue theorem to the expression above yields the same result as in the case of generalized Hubbard-Stratonovich transformation. 
 \section{Asymptotic  Gap Probability and the Microscopic Limit}\label{asymptotics} From a theoretical and a practical point of view it is important to analyze the large $n$ and $p$ limits of the gap probability $E^{(\beta)}_p(t)$. To perform this limit we have to determine a local scale. This is done in \ref{analysismicroscopiclimit}.  In section \ref{statement} we derive an new matrix model with similar asymptotics as the original one,  for $\gamma\in\mathbb{N}$ and $\gamma\in \frac{1}{2} \mathbb N$. Section \ref{asymptoticsSUSY} gives explicit expression for the particular asymptotics, if $\gamma$ is integer.

\subsection{Analysis of the Microscopic Limit}\label{analysismicroscopiclimit}
We now introduce a limit in which the distribution of the smallest eigenvalue becomes universal on a certain local scale. To determine this local scale, we first study the average of the smallest eigenvalue for large $p$ and $n$, but $n-p$ finite. Then we use the level density for an uncorrelated Wishart model, to fix the full $\Lambda$ dependence of the local scale. Employing relation~(\ref{relationpminEgap}), the meanvalue of the smallest eigenvalue is
\begin{equation}
  \eqalign{
 \left<t\right> &= \int\limits_0^{\infty}\dd{t}~t~p^{(\beta)}_{\footnotesize\text{min}}(t)  =  \int\limits_0^{\infty}\dd{t}~E_p^{(\beta)}(t)  }~.
\end{equation}
It is  convenient to use Eq.~(\ref{SuperbosonizationE}) which  yields
\begin{equation}
  \eqalign{
\left<t\right> &=  K_{p\times \bar n} \int\limits_0^{\infty}\dd{t}~    \exp\left(-\tr\frac{\beta t}{2\Lambda}\right)  \int\leb{U} \exp\left(-\tr U \right)\\\fl\quad\quad\quad&\times \Det^{-\kappa}U     \prod_{k=1}^p\Det^{\beta/2} \left(\frac{\beta t}{2}\Id_{2\gamma/\beta}-\Lambda_kU\right)
  ~,} \label{pminmean1}
\end{equation}
where the domain of integration is either $\text{USp}(2\gamma)$ if $\beta=1$ or $\text{U}(\gamma)$ if $\beta=2$. Compared to Eq.~(\ref{SuperbosonizationE}) we rescale the integration variable by $2/\beta \imath t$, \textit{i.e.},  $U\rightarrow 2 U /\beta \imath t $.

The normalization constant is determined by  $\lim_{t\rightarrow 0} E^{(\beta)}(t) =1$ so that $K_{p\times \bar n} = (-1)^p\Det^{\gamma}\Lambda~K_\gamma$, where $K_\gamma$ is finite in the microscopic limit. It is given by
\begin{eqnarray}
K_\gamma &= \int\leb{U} \exp\left(-\tr U \right) \Det^{-\gamma}U~.
\end{eqnarray}
If we combine $\det^{\gamma}\Lambda\det^{-\beta p/2} U$ in Eq.~(\ref{pminmean1}) with the $p$-fold product, we obtain
\begin{equation}
  \eqalign{
%\fl\quad\quad\quad   
\left<t\right> &=  K_{\gamma}\int\limits_0^{\infty}\dd{t}~   \int\leb{U} \exp\left(-\tr U \right)\Det^{-\gamma}U  \\\fl\quad\quad\quad&\times \exp\left(-\tr\frac{\beta t}{2\Lambda}\right) \prod_{k=0}^p\Det^{\beta/2} \left(\frac{\beta t}{2\Lambda_k}U\dagg-\Id_{2\gamma/\beta}\right)~,} \label{pminmean}
\end{equation}
where we cast the full $p$ dependence into the second row of Eq.~(\ref{pminmean}). Thus, we have to study the behavior of this product for large $p$. % To be able to do so  we have to fix the empirical eigenvalues. 

Let the empirical eigenvalues $\Lambda$ be of order $O(1)$ with a finite number of order $O(p^\tau)$ and $\tau>0$, when $p$ tends to infinity and $n-p$ is kept fixed. Under these moderate conditions, we can estimate the invariants of $\Lambda^{-1}$ by
\begin{eqnarray}
0<  \tr \frac{1}{\Lambda^{m}} \leq \frac{p}{\Lambda_{\footnotesize\text{min}}^m},\label{traceinverse}
\end{eqnarray}
where $\Lambda_{\footnotesize\text{min}}$ is the smallest empirical eigenvalue. For an empirical correlation matrix providing such an eigenvalue spectrum, we analyze Eq.~(\ref{pminmean}). If we express the  $p$--fold product in Eq.~(\ref{pminmean}) as a sum in an exponent and expand its argument with respect to Eq.~(\ref{traceinverse}), we find 
\begin{equation} 
\fl\quad\quad\quad\quad\prod_{k=0}^p\Det^{\beta/2} \left(\Id_p-\frac{\beta t}{2\Lambda_k}U\dagg\right)=\exp\left(-\frac{\beta}{2}\sum\limits_{m=0}^\infty\frac{1}{m}\tr\left(\frac{\beta t}{2\Lambda}\right)^m\tr (U\dagg)^m\right)~.\label{detexpansion}
 \end{equation}
Both the exponent in the second row of Eq.~(\ref{pminmean}) and the $p$-fold product in Eq.~(\ref{detexpansion}) depend on the invariants of $t\Lambda^{-1}$ only. But this can be estimated with Eq.~(\ref{traceinverse}), implying that we obtain on the scale $u\sim tp$, 
\begin{equation}
\eqalign{
 &\prod_{k=0}^p\Det^{\beta/2} \left(\Id_p-\frac{\beta u}{2p \Lambda_k}U\dagg\right)  \exp\left(-\tr\frac{\beta u}{2p \Lambda}\right) \\&=  \exp\left(-\tr\frac{\beta u}{2p \Lambda}-\frac{\beta}{2}\tr\frac{\beta u}{2 p\Lambda}\tr U\dagg+O(p^{-1})\right)~.}  \label{largeplimit}
\end{equation}
This holds, because $p/\tr\Lambda^{-1}\rightarrow const.$ for $p \rightarrow \infty$. Hence, the mean value on this local scale $\left<u\right>$ will be a constant in the microscopic limit. 

Suppose we change the scale a bit from $p$ to $p^{1+\alpha}$. Because of a missing damping for $\alpha>0$ and $u\rightarrow \infty$, the integral~(\ref{pminmean}) is divergent. Because of the factor $\exp\left(-O(p^{\alpha})\right)$ in Eq.~(\ref{largeplimit}), the integral of Eq.~(\ref{pminmean}) becomes zero when $\alpha<0 $ and $p\rightarrow \infty$. Thus, we have to study the gap probability on the scale $u\sim pt$.

\subsection{Asymptotic Behavior  of the Gap Probability}\label{statement}
Using the analysis of the previews section, we can perform the microscopic limit of the gap probability in two ways. Either we look at the dual model of Eq.~(\ref{SuperbosonizationE}) or we use orthogonal polynomials. If we want to use the method of orthogonal polynomials, we have to find an uncorrelated Wishart matrix model with the same large--$p$ behavior as Eq.~(\ref{finalESUSY}). While constructing such a model, we derive  the proportionality constant of the local scale from the analysis of the gap probability for an uncorrelated Wishart model. 

We discuss the asymptotics by taking the example of integer $\gamma$, but it can readily be generalized to the case of half-integer $\gamma$. From the results of Ref.~\cite{Forrester1993709}, it turns out that it is appropriate to study the gap probability of an uncorrelated Wishart model with variance $v$ on the local scale $u=4pt/v$. Following the lines of reasoning in section \ref{susymethod}, we can write down a dual matrix model for an uncorrelated Wishart matrix model with variance $ 1/\eta$. It reads
\begin{equation}
  \label{eq:uncorwishartmodel}
\eqalign{\fl\quad\quad\quad\quad \left. E^{(\beta)}_{p}(t)\right|_{\Lambda=\Id_p/\eta}&=K_{p\times\bar n} \exp\left(-\frac{\beta p \eta t}{2}\right) \int\leb{\sigma} \exp\left(-\tr \sigma \right)f_{\bar n ,\beta}(\sigma) \\ 
\fl\quad\quad\quad\quad&\times\Det^{p\beta/2}\left(\frac{\beta t}{2}\Id_{ 2\gamma/\beta}-\frac{1}{\eta}\sigma\right)~,}  
\end{equation}
see Eq.~(\ref{finalESUSY}) for the details. If we choose
\begin{equation}
  \label{eq:definitioneta}
  \eta=\frac{1}{p}\sum_{i=1}^p\frac{1}{\Lambda_i} = \frac{1}{p}\tr\Lambda^{-1}~ 
\end{equation}
and adapt the analysis of the preview section, we obtain, on the local scale $u=\eta 4 tp$, 
\begin{equation}
  \label{GapEquality}
   E^{(\beta)}_{p}\left(\frac{u}{4p\eta}\right) = \left. E^{(\beta)}_{p}\left(\frac{u}{4p\eta}\right)\right|_{\Lambda=\Id_p/\eta} + \mathcal{O}(p^{-1})~,
\end{equation}
\textit{i.e.} both models have the same \textit{microscopic limit}. We summarize out findings in the following statement.
\\\\\textbf{Statement: }\textit{Suppose that $n,p$ tend to infinity, while $n-p$ is kept fixed, the empirical eigenvalues are of the order $\mathcal{O}(1)$ with a finite number of order $\mathcal{O}(p^\tau)$, where $\tau>0$. Under these conditions the dual ordinary and supermatrix models for the gap probability $E^{(\beta)}_{p}(t)$ of Eq.~(\ref{finalESUSY}) and  (\ref{finalhalfpowerexponent}) behave asymptotically like the matrix models}
\begin{eqnarray}
\fl E^{(\beta)}_{p}(t) &\sim\exp\left(-\tr\frac{t \beta}{2\Lambda}\right) \int\leb{\sigma} \Det^{p\beta/2}\left(\frac{t\beta}{2}\Id_{4\gamma/2}-\frac{1}{\eta} \sigma\right) ~ \exp\left(-\tr \sigma\right) ~ f_{\beta}(\sigma)~,\label{asympNormalN}
\end{eqnarray}
\textit{for $\gamma\in \mathbb N$ and for $\gamma \in \mathbb N/2$ as}
\begin{eqnarray}
\fl E^{(1)}_{p}(t) &\sim\exp\left(-\tr\frac{t}{2\Lambda}\right) \int\leb{\mu} \Sdet^{-p/2}\left(\frac{t}{2}\Id_{2\alpha+3}-\frac{1}{\eta} \mu\right) \exp\left(-\Str \mu\right) ~ I_{1}(\mu)~,\label{asympHalfN}
\end{eqnarray}
\textit{where $\eta=\tr \Lambda^{-1}/p$.} \\~\\ We therefore can use the uncorrelated Wishart model to study the microscopic limit. This has already been studied  in the context of sample correlation matrices, QCD and telecommunication, \textit{c.f.} Refs.~\cite{Edelman1991,GuhrWettigWilke,Forrester1993709,DamgaardNishigaki,KatzavPerez}. Instead of employing these results we work them out using the expressions obtained from the dual model.
\subsection{Asymptotics Using the Dual Model}\label{asymptoticsSUSY}
We consider the asymptotics in view of the microscopic limit. We only look at $\gamma \in \mathbb N$ and utilize the large-$p$ behavior of $E^{(\beta)}_{p}(t)$ given in the statement above. Because of its simple structure we study it using expressions of section \ref{exactSB}.

Since the determinant and the Pfaffian kernel of Eq.~(\ref{startpointasymptoticscomplex}) and (\ref{startpointasymptoticsreal}) are of the same kind we  analyze them together. They are given, after appropriate redefinitions, by integrals of the form
\begin{eqnarray}
\Omega_m(t;\Lambda) &= \oint\limits_{|z|=1}\dd{z}~z^{m-p}~e^{z} \prod_k^p \left(t + \Lambda_kz\right)~,
\end{eqnarray}
where $m$ is an arbitrary integer, that is of the order $\mathcal{O}(1)$ in the microscopic limit. To study the asymptotics we use the arguments of section \ref{analysismicroscopiclimit} and approximate the $p$-fold product by
\begin{eqnarray}
  \prod_k^p \left(t + \Lambda_kz\right) \approx \Det \Lambda ~z^{p}~\exp\left(\frac{1}{z}\tr\frac{t}{\Lambda}\right)~.\label{eq:oneDapproximation}
\end{eqnarray}
As we will see, the $p$ as well as the $n$-dependence of our expression for the gap probability disappears and we can perform  $p$ limit without struggling  with cumbersome expressions. Substituting the approximation~(\ref{eq:oneDapproximation}) into $\Omega_m(t;\Lambda)$ yields 
\begin{eqnarray}
\Omega_m(t;\Lambda) &\approx \Det\Lambda \oint\limits_{|z|=1}\dd{z}~z^{m}~\exp\left(z+\frac{1}{z}\tr\frac{t}{\Lambda}\right)~ .\label{Kasymptotic}
\end{eqnarray}
The closed contour integral is, up to a constant, the definition of the modified Bessel function of first kind $\text{I}_m$. To see this, we rescale the integration measure by the square root of  $t \tr \Lambda^{-1}$ and use the expansion  \cite{AbramowitzStegun}
\begin{eqnarray}
  \exp\left(w\left(z+\frac{1}{z}\right)\right) &= \sum^\infty_{k=-\infty}z^k \text{I}_{k}(2w)~,\quad \forall z\neq 0~.
\end{eqnarray}
Evaluation of the remaining contour integral then projects out only one of the terms in this Laurent series such that $\Omega_m(t;\Lambda)$ is approximately given by
\begin{eqnarray}
 \Omega_m(t;\Lambda) &\approx 2\pi\imath~\Det\Lambda ~ \sqrt{\tr\frac{t}{\Lambda}}^{m+1} ~\text{I}_{m+1}\left(2\sqrt{\tr\frac{t}{ \Lambda}}\right) ~.
\end{eqnarray}
If we substitute this asymptotic expression into Eq.~(\ref{startpointasymptoticscomplex}) and Eq.~(\ref{startpointasymptoticsreal}) with $m=i+j-2-\gamma$ and $m=i+j-3-2\gamma$, respectively, and go on the local scale, we obtain for the microscopic limit of the gap probability,
\begin{eqnarray}
 \mathcal E^{(\beta)}(u) &= \exp\left(- \frac{\beta u}{8}\right)
    \Det^{\beta/2}\left[\tilde q_{ij}~L^{(0)}_{ij}(u)\right] 
~,  \label{asympbeta2final}
\end{eqnarray}
where
\begin{eqnarray}
 L^{(l)}_{ij}(u) &= \sqrt{u/4}^{i+j -\kappa'} \text{I}_{\kappa'+\delta_{i-l,0} -i-j}\left(\sqrt{u}\right)~.
\end{eqnarray}
We use the upper index $(l)$ for later purpose and we also introduce $\kappa'=2(\gamma+1)/\beta$ and $\tilde q_{ij}=(j-i)$ for $\beta=1$, $\tilde q_{ij}=(-1)^{i+1}$ for $\beta=2$. The normalization follows from the small $z$ expansion of the modified Bessel function  \cite{AbramowitzStegun},
\begin{eqnarray*}
  I_m(2z) &\sim \frac{z^{m}}{\Gamma(m+1)}~,\quad \forall m\geq 0,1,2,\dots 
\end{eqnarray*}
and  $\mathcal E^{(\beta)}(u)\rightarrow 1$ for  $u \rightarrow 0$. The normalization turns out to be  $\Det^{-\gamma}\Lambda$ such that it cancels the factor $\det \Lambda$ in Eq.~(\ref{Kasymptotic}).
\section{Distribution of the Smallest Eigenvalue}\label{Mindistribution}
The distribution of the smallest eigenvalue $\mathcal P^{(\beta)}_{\footnotesize \text{min}}(t)$  and the gap probability $E^{(\beta)}_{p}(t)$ are related by Eq.~(\ref{relationpminEgap}). We compute this probability distribution for $\gamma\in \mathbb N$ and both values of $\beta$. Since we have exact and asymptotic results, both are considered. Although calculations are similar we consider the real and the complex case separately. We start with $\beta=2$ in section \ref{pmincomplex} and go over to $\beta=1$ in section \ref{pminreal}.
\subsection{Complex Case}\label{pmincomplex}
We start with the exact results in section \ref{pmincomplexexact} and compute the asymptotic distribution of the smallest eigenvalue in section \ref{pmincomplexasympt}.
\subsubsection{Exact Results}\label{pmincomplexexact}
The exact result for the gap probability $E^{(2)}_{p}(t)$ was computed in section \ref{exactResults} and can be found in Eq.~(\ref{Ebeta2SuperB}). Differentiation of Eq.~(\ref{Ebeta2SuperB}) with respect to $t$ yields the distribution of smallest eigenvalue 
\begin{eqnarray}
\fl\quad\quad\quad\quad  \mathcal P^{(2)}_{\footnotesize\text{min}}(t) &= \tr\frac{1}{\Lambda}~E^{(2)}_{p}(t) - \frac{\exp\left(\displaystyle-\tr\frac{t}{\Lambda}\right)}{\Det^\gamma\Lambda }\sum_{l=1}^\gamma\Det\left[\left. G^{(l)}_{ij}(t) \right|_{i,j=1,\dots, \gamma}\right] , \label{exactpminbeta2}
\end{eqnarray}
where we  defined
\begin{eqnarray}
\fl\quad\quad\quad\quad  G^{(l)}_{ij}(t) &=\Theta(\alpha_{p,2})(-1)^{i+1} \left\{\begin{array}{l@{}l}\displaystyle
\sum_{k=0}^{\text{min}(p,\alpha_{p,2})} \frac{e_k(\Lambda)~t^{p-k}}{(\alpha_{p,2}-k)!} &,l\neq i\\\displaystyle
\sum_{k=0}^{\text{min}(p-1,\alpha_{p,2})} \frac{e_k(\Lambda)~ (p-k)~t^{p-k-1} }{(\alpha_{p,2}-k)!}&,l=i
\end{array}\right.~.
\end{eqnarray}
The normalization of the distribution of the smallest eigenvalue $\mathcal P^{(2)}_{\footnotesize\text{min}}(t)$ in Eq.~(\ref{exactpminbeta2}) follows from the normalization of  $E^{(2)}_{p}(t)$.
\subsubsection{Microscopic Limit}\label{pmincomplexasympt}
The asymptotic expression for the distribution of the smallest eigenvalue is given by a rescaled version of Eq.~(\ref{relationpminEgap}). We have to differentiate the asymptotic expression of the kernel $\Omega_m(t;\Lambda)$, \textit{i.e.} Eq. (\ref{Kasymptotic}), with respect to $t$. Differentiation yields
\begin{equation}
\eqalign{
\fl\quad\quad\quad\quad\frac{\text{d}}{\text{d}t}  \Omega_m(t;\Lambda) &\approx \Det\Lambda \tr \frac{1}{\Lambda} \oint\limits_{|z|=1}\dd{z}~z^{m-1}~\exp\left(z+\frac{1}{z}\tr\frac{t}{\Lambda}\right)\nn\\&=\tr\frac{2\pi \imath}{\Lambda} \Det\Lambda~ \sqrt{\tr\frac{t}{\Lambda}}^{m} ~\text{I}_{m}\left(2\sqrt{\tr\frac{t}{ \Lambda}}\right)~,} \label{dervativeWRTtofK}
\end{equation}
such that
\begin{eqnarray}
\fl\quad\quad\quad\quad    \wp^{(2)}_{\footnotesize\text{min}}(u) &= \frac{1}{4}\mathcal E^{(2)}(u) -\frac{1}{4\sqrt{u}} \exp\left(-\frac{u}{4}\right)\sum_{l=1}^\gamma\Det
\left[\left.\tilde q_{ij}~L^{(l)}_{ij}(u) \right|_{i,j=1,\dots, \gamma}\right]~. \label{pminasymptoticbeta2}
\end{eqnarray}
is the microscopic limit of the distribution of the smallest eigenvalue on the local scale $u=4p\eta t$ in the complex case.
\subsection{Real Case}\label{pminreal}
The analysis of the distribution of the smallest eigenvalue for $\beta=1$ is similar to the previews section. The main difference is the appearance of a Pfaffian instead of a determinant. We give  exact results in section \ref{pminrealexact} and compute the asymptotics in section \ref{pminrealasymp}.
\subsubsection{Exact Results}\label{pminrealexact}
To analyze the structure of the distribution of the smallest eigenvalue $\mathcal P^{(1)}_{\footnotesize \text{min}}(t)$ we consider Eq.~(\ref{FinalExactEbeta1}) and apply Eq.~(\ref{relationpminEgap}). It yields 
\begin{equation}
\fl\quad  \mathcal P^{(1)}_{\footnotesize \text{min}}(t) = \tr\frac{1}{2\Lambda}E^{(1)}_{p}(t) - \frac{1}{2}\frac{\exp\left(-\displaystyle \tr\frac{t}{2\Lambda}\right)\sum\limits_{l=1}^{2\gamma}\Det\left[ ~G^{(l)}_{ij}(t)\right]}{\Det^{\gamma}\Lambda \Pf\left[ G^{(0)}_{ij}(t)\right]}~ ,  \label{pminexactbeta1}
\end{equation}
where we defined, for $\beta=1$,
\begin{eqnarray}
\fl\quad  G^{(l)}_{ij}(t) &= (j-i)(-1)^{i+j}\Theta(\alpha_{p,1})  \left\{\begin{array}{l@{}l}\displaystyle
\sum^{\text{min}(p,\alpha_{p,1})}_{k=0}\frac{e_k(\Lambda)~t^{p-k}}{(\alpha_{p,1}-k)!}~&,l\neq i\\\displaystyle
 \sum^{\text{min}(p-1,\alpha_{p,1})}_{k=0}\frac{e_k(\Lambda)~(p-k)~t^{p-k-1}}{(\alpha_{p,1}-k)!}~&,l=i
\end{array}\right.~.
\end{eqnarray}
We have to stress that Eq.~(\ref{pminexactbeta1}) is apart from the exponential a polynomial in $t$. This is caused by the fact that we are differentiating a polynomial. To derive this expression we use that $\Pf ~M = \sqrt{\det ~M}$, which is true for every antisymmetric even dimensional matrix $M$.
\subsubsection{Microscopic Limit}\label{pminrealasymp}
Asymptotic expression for the real ensemble are derived from  Eq.~(\ref{relationpminEgap}) and Eq.~(\ref{dervativeWRTtofK}). We find 
\begin{eqnarray}
   \wp^{(1)}_{\footnotesize \text{min}}(u) = \frac{1}{8}\mathcal{E}^{(1)}(u) - \frac{\exp\left(\displaystyle-\frac{u}{8}\right)\sum_{l=1}^{2\gamma}\Det\left[L^{(l)}_{ij}(u)\right] }{8\sqrt{u}~\Pf\left[L_{ij}^{(0)}(u)\right]}~.  \label{pminasymptoticbeta1}
\end{eqnarray}
for the microscopic limit of the distribution of the smallest eigenvalue in the real case.
\section{Numerical Simulations}\label{numerics}
\begin{figure}[ht]
  \centering
  \includegraphics[width=\textwidth]{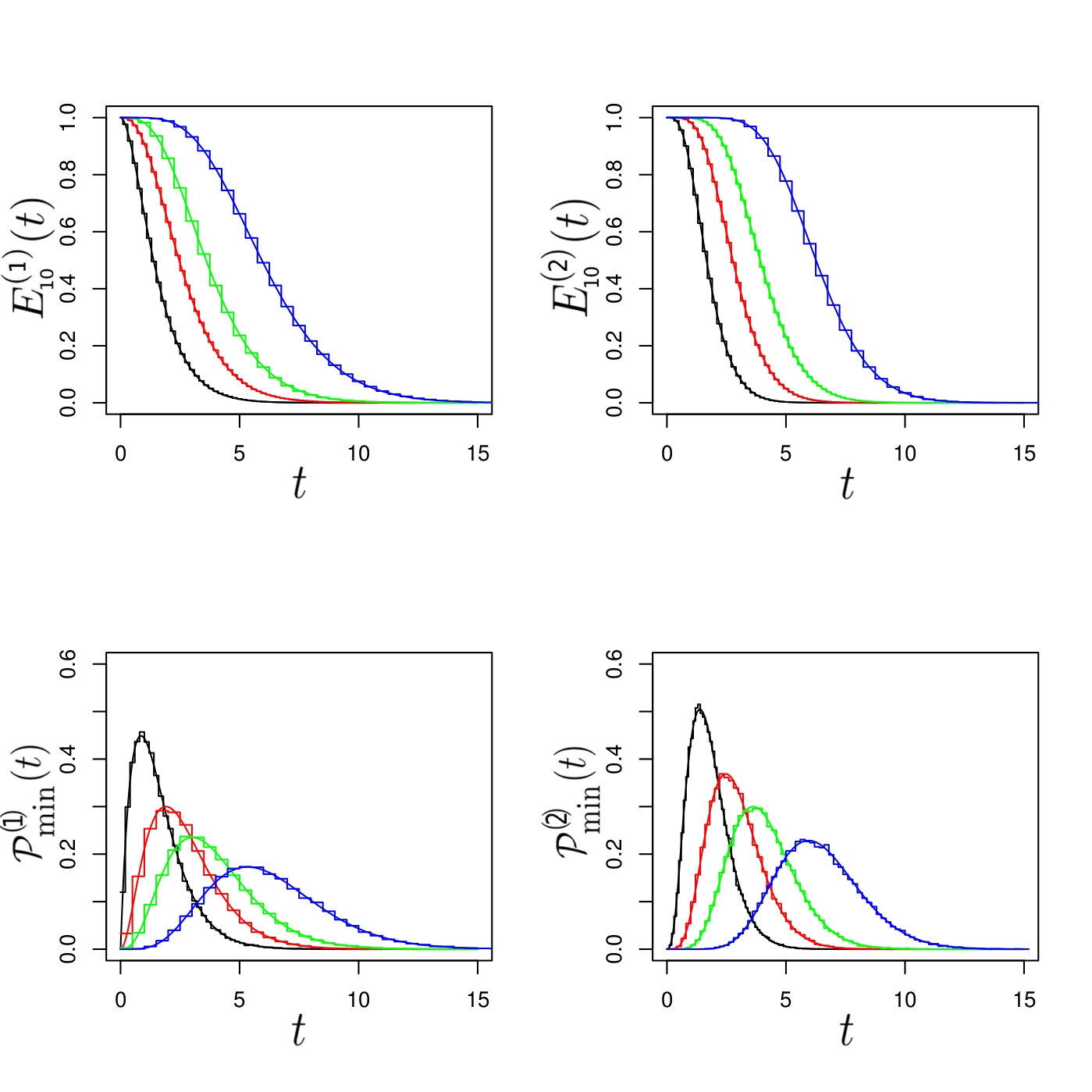}
  \caption{(color online) The upper two figures show $E^{(\beta)}_{p}(t)$ and the lower $\mathcal P^{(\beta)}_{\footnotesize \text{min}}(t)$ for fixed $p=10$, where $\Lambda_k=0.6,1.2,6.7,9.3,10.5,15.5,17.2,20.25,30.1,35.4$ and $n=13,15,17,21$. The left figures correspond to the real ($\beta=1$) and the right to the complex ensemble ($\beta=2$) . The straight lines are the analytic and the step functions are the numeric results. We use 50,000 samples drawn from a Gaussian distribution.}
  \label{fig}
\end{figure}

Although our results are exact we compare them to numerical simulations for illustrating purpose and to confirm the validity and correctness of our final expressions. We implement the formulas into the computer code R  \cite{RMan} and generate 50000 correlated random Wishart matrices drawn from the distribution of Eq.~(\ref{WRMdistribution}) for both ensembles, the real and the complex. From the analysis of section \ref{exactResults} and section \ref{Mindistribution} we known that the rectangularity governs the dimension of the dual matrix models. Thus, we carry out the simulations for four different rectangularities. The results are shown in Fig. \ref{fig}. As eigenvalues of the sample correlation matrix we choose $\Lambda_k=0.6,1.2,6.7,9.3,10.5,15.5,17.2,20.25,30.1,35.4$ for both, the real and the complex ensemble. The figures show perfect agreement of the analytic and the numerical results. 

\begin{figure}[ht]
  \centering
  \includegraphics[width=0.7\textwidth]{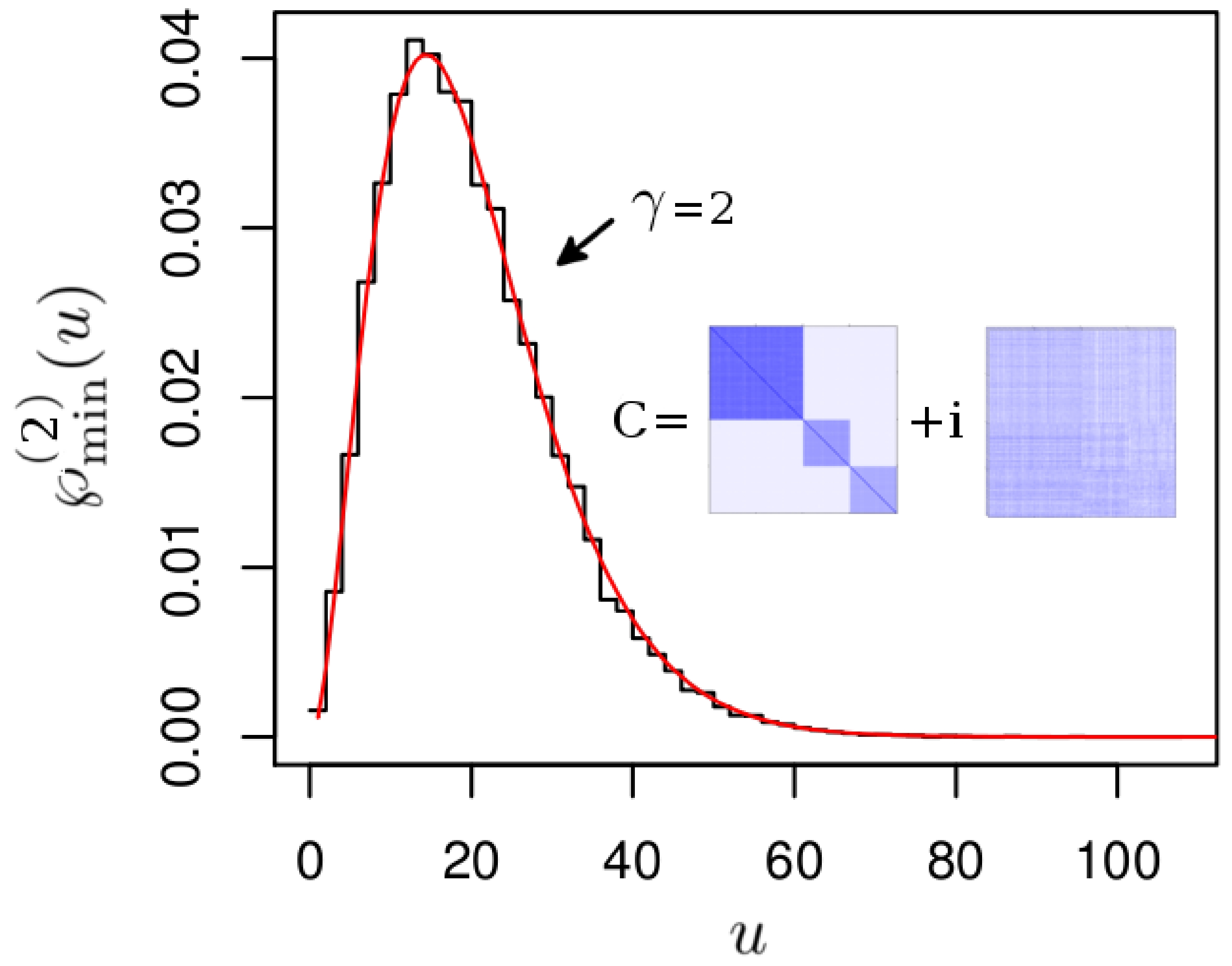}
  \caption{(color online) The microscopic limit of the distribution of the smallest eigenvalue for $\beta=2$, \textit{i.e.}, $\wp^{(2)}_{\footnotesize\text{min}}(u)$. The straight lines are the analytic and the step functions are the numeric results generated from 30,000 samples of $200\times 202$-dimensional complex correlated Wishart matrices.}
  \label{fig2}
\end{figure}

To emphasize our findings for the microscopic limit of the distribution of the smallest eigenvalue $\wp^{(\beta)}_{\footnotesize\text{min}}(u)$, we produce a non-trivial empirical correlation matrix and generate 30,000  samples of complex correlated $200\times202$-dimensional Wishart matrices. The structure of empirical correlation matrix $C$ is indicated in Fig. \ref{fig2}. We compare our analytic findings for the distribution of the smallest eigenvalue on the local scale with the numerical simulations. Once more, we obtain a perfect agreement of the simulations and our analytic results shown in Fig.~\ref{fig2}.
\section{Conclusion}\label{summary}
Our results have three aspects, a conceptual, a practical and a universal one. On the conceptual side we discuss mutual dualities of matrix models which then helped as to derive exact formulas of practical relevance. Moreover, we identify a new universality for all real and complex correlated Wishart ensembles.

On the conceptual level, we found infinitely many dualities between statistical quantities. The infinite number of possibilities reflects the freedom in choosing that dimension of the matrices which corresponds to the number of time steps. In turn, each of these models has a dual model in superspace with, in general, different bosonic and fermionic dimensions. Our most important result is the discovery of the duality between the $W$ and the $\overline{W}$ matrix models, because the bosonic dimension of the supersymmetric dual is zero and it therefore leads to a model which collapses to an ordinary, invariant matrix model. 

Although we used only the small-$\sigma$ model in the main text, we discussed the other dual models as well to emphasize the significant simplification that the small-$\sigma$ model entails. If we used, for instance, the large-$\sigma$ model to study the smallest eigenvalue, we  would have to compute the supersymmetric Itzykson-Zuber integral as well as the Efetov-Wegner term for $\beta=1,2$. However, if $\beta=1$, these non-trivial objects are known in a few situations only.

The exact formulas constitute the major part of this contribution. We have shown that it is possible, even for $\beta=1$, to obtain a determinant respectively a Pfaffian structure for the distribution of the smallest eigenvalue and a related gap probability. Up to an exponential, the expressions for  $E^{(\beta)}_p(t)$ and $\mathcal P_{\footnotesize\text{min}}^{(\beta)}(t)$ are finite polynomials in $t$ and the empirical eigenvalues $\Lambda_k$. The compact and easy-to-use structure of  our results serve as a  starting point for further analysis and applications, because the formulas can be evaluated, even for large matrix dimensions, much faster and with a higher precision than numerical simulations.

The difficulty for $n-p$ even is caused by a characteristic polynomial with half-integer power in Eq.~(\ref{finalE}). Nonetheless, we were able to express, even in this case, the gap probability as a full supermatrix model which is invariant under the action of a certain symmetry group. But diagonalization leads to an Efetov-Wegner or Rothstein term  \cite{Rothstein}. This term is highly non-trivial and for $\beta=1$ yet unknown. We leave the computation of the remaining supermatrix integral to future work.

The local, microscopic scale that we identified leads to an universal distribution of the smallest eigenvalue for arbitrary correlation structures. The controlling parameters are the size of the matrix, the symmetry class and the empirical correlation matrix $C$. We were able to show that in the microscopic limit the gap probability as well as the distribution of the smallest eigenvalue become independent of the empirical correlation matrix. This means that the statistics at the lower edge of the spectrum on a local scale is governed by the universal fluctuations. 
\section*{Acknowledgment}
We thank Rudi Sch\"afer for fruitful discussions on applications of our results and Santosh Kumar and Mario Kieburg for discussions and many useful comments.  We acknowledge support from the German Research Council (DFG) via the Sonderforschungsbereich Transregio 12, ``Symmetries and Universality in Mesoscopic Systems''.

\appendix
\section{Applications of the Smallest Eigenvalue}\label{appofsmallestEigenvalue}
The aim of this section is to illustrate applications of the smallest eigenvalue in different areas of multivariate statistics. We concentrate on examples in high dimensional inference as well as applications in numerical analysis, telecommunication and portfolio theory.

 Linear or Gaussian discriminant analysis is a method which is used to classify measurements in data analysis. Suppose we have $k$ observations of  normal distributed $p$-dimensional variates $X_i$, with mean zero and unit variance. We want to classify the data into two classes. These classes correspond to ensembles drawn from normal distributions with the correlation matrices $C_y$, where $y=0,1$. Linear discriminant analysis is a rule deciding to which class an observation most likely belongs~\cite{LarryWasserman}. For a particular observation $X_i$, one has to evaluate
\begin{eqnarray}
\mu(X_i)=  \left(X_i\dagg C_0^{-1}X_i\right)^2 + \log\frac{\Det C_0}{\Det C_1} +2\log\frac{\pi_1}{\pi_0}-  \left(X_i\dagg C_1^{-1}X_i\right)^2~,  \label{LDAthreshold}
\end{eqnarray}
$\pi_j$ are free controlling parameters satisfying $\pi_0+\pi_1=1$, to decide to which class it belongs. These are known as prior probability of the class $i$. In applications they are determined using training sets. If the linear function $\mu(X_i)$ is below zero, $X_i$ belongs to class $y=1$, otherwise $X_i$ belongs to $y=0$. 

Assuming we have a set of $p$-variates, where $p$ is large,  it is consistent with empirical observations to presume that the ``real'' statistics lie approximately on a submanifold in $\IR^p$. If it is described by linear equations, it is a flat plane in $\IR^p$. Linear principle component analysis is a method to determine a linear plane in the space of $p$-variates that is close to all observations \cite{Gnanadesikan}. The best fitting plane, closest to all measurements, is described by the eigenvector corresponding to the smallest eigenvalue of the correlation matrix $C$ of the system.

By definition single statistical outliers lie far from the center of observation. The distance is measured, \textit{e.g.}, using Mahalanobis distance
\begin{eqnarray}
  \label{mahalanobisdist}
  \Omega_{\nu,C}(X,Y) = \sqrt{(X-\nu)\dagg C^{-1} (Y-\nu)}~,
\end{eqnarray}
where $\nu$ and $C$ are the sample mean value  and correlation matrix, respectively. It is maximized by the eigenvector corresponding to the smallest eigenvalue of $C$ ~\cite{BarnettLewis}.

Another example of higher dimensional inference are the statistics of the condition number of a random matrix \cite{Edelmann1988,Edelman1991,Edelman1992}. It is crucial to know the distribution of the condition number to study the statistics of numerical errors in data analysis. Because the precision of a numerical solution to a linear equation including a large random matrix is bounded by the condition number $\kappa$. If the $\textit{l}^2$-norm is considered, it can be shown that this number is given by
\begin{equation}
\kappa(A) = \left|\frac{\lambda_{\footnotesize\text{max}}(A)}{\lambda_{\footnotesize\text{min}}(A)}\right|~,
\end{equation}
where $\lambda_{\footnotesize\text{min/max}}(A)$ denotes the smallest and the largest singular value of $A$. It is the square root of the smallest, respectively, largest eigenvalue of $AA\dagg$.

In wireless telecommunication Wishart matrices are used to model Multi--Input--Multi--Output channel matrices of antenna arrays~\cite{FoschiniGans}. The model is valid under the assumption of a narrow bandwidth and slow environmental fading~\cite{TulinoVerdu}. If Rayleigh fading is present, the distribution of the uncorrelated complex Wishart matrix is consistent with the empirical observations~\cite{digitalcommunication}. Moreover, compact antenna architectures in transmitting and receiving antenna lead to feedback, which induces row and  column-wise correlation in the channel matrix~\cite{ChenTseKahnValenzuela,UpamanyuMadhow}. The case considered here corresponds to feedback in the receiver system only. 

In digital communication the signals are transmitted using symbols from a finite symbol set. The purpose of the receiver architecture is to estimate a symbol from a received signal. This estimate has an error which is bounded by the smallest eigenvalue of the channel matrix \cite{Burel02statisticalanalysis}. To optimize certain symbol identification algorithms, it is therefore gainful to know the statistics of the smallest eigenvalue.

 A last example comes from finance mathematics. The optimal portfolio depends linearly on the inverse correlation matrix \cite{EltonGruber,Markowitz}. Thus, it is governed by the largest eigenvalue of $C^{-1}$, which is the smallest eigenvalue of $C$.

\section{Supersymmetric Representation for the Generating Function}\label{generalizedHS}
By adapting the work of \cite{Guhr2006I,KieburgGroenqvistGuhr}, we sketch the application of the generalized Hubbard-Stratonovich transformation to the present case. The results summarized here are a generalization of the results obtained in Ref.~\cite{Recheretal}. It is illustrated by taking the example of $\beta=2$, but it can readily be extended to $\beta=1$. A more general and detailed analysis will given elsewhere \cite{GuhrWirtzIII}. 

Let $W$ be a complex $p\times n$ matrix, we introduce the generating function 
\begin{equation}
\fl\quad\quad\eqalign{
Z^{(2)}_{k_1,k_2}(\kappa) &= \frac{1}{\mathcal N} \int \leb{W} \exp\left(-\tr WW\dagg \Lambda^{-1}\right) \frac{\prod_{i=1}^{k_2}\Det\left(WW\dagg - \kappa_{i,2}\Id_p\right)}{\prod_{j=1}^{k_1}\Det\left(WW\dagg - \kappa_{j,1}\Id_p\right)}\\&=\frac{1}{\mathcal N} \int \leb{W}\exp\left(-\tr WW\dagg\right) \frac{\prod_{i=1}^{k_2}\Det\left(WW\dagg\Lambda - \kappa_{i,2}\Id_p\right)}{\prod_{j=1}^{k_1}\Det\left(WW\dagg\Lambda - \kappa_{j,1}\Id_p\right)}~,} \label{definitionZ}
\end{equation}
where $\mathcal N$ is determined by $\lim_{\kappa\rightarrow0}Z^{(2)}_{k_1,k_2}(\kappa) =1$ and the $\kappa_{i,1}$ are chosen such that the integral of Eq.~(\ref{definitionZ}) exists.  The ratio of determinants in Eq.~(\ref{definitionZ}) can be written in form of Gaussian integrals. The determinants in the denominator as integral over $k_1$ complex $p$-dimensional vectors $z_i$, $i=1,\dots,k_1$, and those in the numerator as integrals over $k_2$ complex $p$-dimensional vectors with Grassmannian  entries $\zeta_j$, $j=1,\dots,k_2$,
\begin{equation}
\eqalign{
\fl\quad\quad  \frac{\prod_{i=1}^{k_2}\Det\left(WW\dagg\Lambda - \kappa_{i,2}\Id_p\right)}{\prod_{j=1}^{k_1}\Det\left(WW\dagg\Lambda - \kappa_{j,1}\Id_p\right)} &\sim \int \prod_{i=1}^{k_1}\leb{z_i}\exp\left(\imath z_i\dagg\left(WW\dagg\Lambda - \kappa_{i,1}\right)z_i\right)\\&\times\int\prod_{j=1}^{k_2}\leb{\zeta_j} \exp\left(\imath \zeta_j\dagg\left(WW\dagg\Lambda - \kappa_{j,1}\right)\zeta_j\right)~.}
\label{detratio}
\end{equation}
For the details on integration over Grassmannian variables we refer to Ref.~\cite{Berezin}. If we introduce the matrix 
\begin{eqnarray}
  \label{definitionA}
  A =\left[ \begin{array}{cccccc} z_{1} &\dots &z_{k_1}& \zeta_{1} &\dots&\zeta_{k_2}
  \end{array}\right]~,
\end{eqnarray}
and its super Hermitian conjugate $A\dagg$, the right hand side of Eq.~(\ref{detratio}) can be cast into the from
\begin{equation}
\fl\quad  \frac{\prod_{i=1}^{k_2}\Det\left(WW\dagg\Lambda - \kappa_{i,2}\Id_p\right)}{\prod_{j=1}^{k_1}\Det\left(WW\dagg\Lambda - \kappa_{j,1}\Id_p\right)} \sim \int \leb{A}\exp\left(-\imath \Str A\dagg A \kappa +\imath\tr AA\dagg WW\dagg\Lambda\right)~,
\label{detratiowithW}
\end{equation}
where we used $\kappa=\text{diag}(\kappa_{1,1},\dots,\kappa_{k_1,1},\kappa_{1,2},\dots,\kappa_{k_2,2})$ and defined the measure $\leb{A}=\prod_{i=1}^{k_1}\leb{z_i}\prod_{j=1}^{k_2}\leb{\zeta_j}$. 

If we substitute Eq.~(\ref{detratiowithW}) into Eq.~(\ref{definitionZ}) and exchange the order of $\leb{W}$ and $\leb{A}$, we find the characteristic function with respect to Eq.~(\ref{WRMdistribution}). The characteristic function $\Phi(K)$, where $K$ is an arbitrary Hermitian matrix, is defined as the ensemble average of $\exp\left(\imath\tr KWW\dagg\right)$,
\begin{eqnarray}
  \fl\quad\quad\Phi(K) &= \int \leb{W} \exp\left(-\tr WW\dagg\right)\exp\left(\imath\tr KWW\dagg\right)\sim \Det^{n}\left(\Id_p-\imath K\right)~.
\label{definitionPhiK}
\end{eqnarray}
From Eq.~(\ref{definitionPhiK}) it turns out that the characteristic function is invariant under the adjoint action of the unitary group, \textit{i.e.} $K\rightarrow UKU\dagg$, where $U\in \text{U}(p)$. Replacing it in the generating function by Eq. (\ref{definitionPhiK}) yields 
\begin{eqnarray}
Z^{(2)}_{k_1,k_2}(\kappa) &= \frac{1}{\mathcal N} \int \leb{A}\exp\left(-\imath \Str A\dagg A \kappa \right) \Det^{n}\left(\Id_p-\imath \Lambda AA\dagg\right)~.
\label{Arepresentation}
\end{eqnarray}
The integrand above depends partially on the invariants $\tr \left(\Lambda AA\dagg\right)^m$, $m\in\mathbb N$, but these correspond to invariants of the $(k_1|k_2)$-dimensional supermatrix $A\dagg\Lambda A$~\cite{Recheretal}, \textit{i.e.}
\begin{equation}
 \tr\left(A\dagg \Lambda A\right)^m=  \Str\left(A\dagg \Lambda A\right)^m~.
\end{equation}
We substitute in the determinant in Eq.~(\ref{Arepresentation}), for the invariants of $\Lambda A A\dagg$, the invariants of $A\dagg \Lambda A$ and obtain
\begin{equation}
 \Det^{n}\left(\Id_p-\imath \Lambda AA\dagg\right) = \Sdet^{n}\left(\Id_{k_1+k_2}-\imath A\dagg \Lambda A\right).
\end{equation}
In a final step to construct the supermatrix model, we replace $A\dagg\Lambda A$ in the superdeterminant above by a supermatrix with the same symmetries using an integral over the supersymmetric delta function~\cite{LehmannSaherSokolovSommers}
\begin{eqnarray}
\delta(\sigma - A\dagg\Lambda A) &\sim \int \leb{\varrho} \exp\left(-\imath\Str\varrho(\sigma-A\dagg\Lambda A)\right)~.
\label{definitionDelta}
  \end{eqnarray}
From Eq.~(\ref{definitionA}) it turns out that the matrix $A\dagg\Lambda A$ is Hermitian. Hence, we have to integrate over the set of $(k_1|k_2)$-dimensional Hermitian supermatrices $\sigma,\varrho$ parametrized by~\cite{GuhrSUSY}
\begin{eqnarray}
 \sigma &= \left[\begin{array}{cc} \sigma_{BB} & \eta \\ -\eta\dagg & \imath\sigma_{FF}\end{array}\right] \quad\text{and} \quad\varrho &= \left[\begin{array}{cc} \varrho_{BB} & \vartheta \\ -\vartheta\dagg & \imath\varrho_{FF}\end{array}\right],
\end{eqnarray}
where $\sigma_{BB},\varrho_{BB}$ and $\sigma_{FF},\varrho_{FF}$ are ordinary Hermitian $k_1\times k_1$ respectively $k_2\times k_2$ matrices and $\eta,\vartheta$ are  $k_1\times k_2$ rectangular matrix with Grassmannian entries. The factor of $\imath$ in front of $\sigma_{FF},\varrho_{FF}$ ensures the convergence of integrals. As measure on the space of $(k_1|k_2)$-dimensional Hermitian supermatrices we use the usual flat one, 
\begin{eqnarray}
  \leb{\sigma} &= \leb{\sigma_{BB}}\leb{\sigma_{FF}}\leb{\eta} ~,
\end{eqnarray}
consisting of the product of all independent differentials of $\sigma_{BB}$ and $\sigma_{FF}$ as well as $\leb{\eta} = \prod_{i=1,j=1}^{k_1,k_2} \dd{\eta^*_{ij}}\dd{\eta_{ij}}$. The same is true for the $\varrho$ integration.

Utilizing the supersymmetric delta function~(\ref{definitionDelta}), we represent the superdeterminant as double integral over two Hermitian supermatrices \cite{Guhr2006I,KieburgGroenqvistGuhr},
\begin{equation}
\eqalign{
   \Sdet^{n}\left(\Id_{k_1+k_2}-\imath A\dagg \Lambda A\right) &\sim \int\leb{\varrho}\leb{\sigma}\Sdet^{n}\left(\Id_{k_1+k_2}-\imath \sigma\right)\\&\times\exp\left(-\imath\Str\varrho(\sigma-A\dagg\Lambda A)\right)~.}
\label{superrepPhi}
  \end{equation}
Inserting Eq.~(\ref{superrepPhi}) into Eq.~(\ref{Arepresentation}) and exchanging the $A$ and the $\varrho$, $\sigma$ integrals yields 
\begin{equation}
\eqalign{
Z^{(2)}_{k_1,k_2}(\kappa) &= \frac{1}{\mathcal N}\int\leb{\varrho}\leb{\sigma}\Sdet^{n}\left(\Id_{k_1+k_2}-\imath \sigma\right) \exp\left(-\imath\Str\varrho\sigma\right)\\&\times \int \leb{A}\exp\left(\imath \Str \left(\varrho A\dagg\Lambda A-A\dagg A \kappa\right) \right)~,}
\label{sigmarhoArep}
\end{equation}
The $A$ integral simplifies to a standard Gaussian one, which is known in the literature \cite{Verbaarschotetsal}. We introduce the supersymmetric probability distribution as the Fourier back transformed of the characteristic function,
\begin{eqnarray}
Q(\varrho) &= \int\leb{\sigma}\Sdet^{n}\left(\Id_{k_1+k_2}-\imath \sigma\right) \exp\left(-\imath\Str\varrho\sigma\right).
\label{definitionQ}
\end{eqnarray}
If we shift $\sigma$ by $-\imath\Id_{4\gamma/\beta}$, this does not effect the domain of integration \cite{Guhr2006I,KieburgGroenqvistGuhr} and yields  
\begin{eqnarray}
  \label{eq:QandIS}
  Q(\varrho) &= \exp\left(-\Str \varrho \right)I_{2}(\varrho),
\end{eqnarray}
where
\begin{eqnarray}
I_{\beta}(\varrho) &= \int \leb{\nu} ~\Sdet^{-\bar n\beta/2}\nu~\exp\left(-\imath \Str \mu \nu\right)  \label{Ingham-Siegel} 
\end{eqnarray}
is known as the supersymmetric Ingham-Siegel Integral. It is a distribution on the space of Hermitian supermatrices, invariant under the adjoint action of $\text{U}(k_1|k_2)$ and an analytic solution in eigenvalue representation is known in the literature \cite{Guhr2006I,KieburgGroenqvistGuhr}.

If we subsutitute this into the generating function, we obtain its supersymmetric representation 
\begin{equation}
\fl\quad\quad\quad Z^{(2)}_{k_1,k_2}(\kappa) = \frac{1}{\mathcal N}\int\leb{\varrho}\exp\left(-\Str \varrho \right)I_{2}(\varrho)\prod_{k=1}^p\Sdet^{-1}\left(\Lambda_k\varrho-\kappa\right)~.
\label{finalZSUSY}
\end{equation}
The structure of the supersymmetric representation for $\beta=1$ is similar to Eq.~(\ref{finalZSUSY}), but the integration domain and the matrix $\kappa$ are different.

% we discuss a supersymmetric dual of Exq.~(\ref{finalE}). The characteristic function with respect to the distribution~(\ref{WRMdistribution}) is \cite{muirhead}
% \begin{eqnarray}
%   \label{characteristicfunctionWRM}
%   \Phi(K|\Lambda) &= \Det^{-n}\left(\Id_p - \imath \Lambda K\right)~.
% \end{eqnarray}
% Inserting this into Eq.~(\ref{definitionQ}) we find that $Q(\varrho)$ is given by
% \begin{eqnarray}
%   \label{shiftedIngham-Siegel}
% Q(\varrho)  &= \int \leb{\sigma} ~\Sdet^{-\bar n\beta/2}\left(\Id_{4\gamma/\beta}-\imath\sigma\right)\exp\left(-\imath \Str \sigma\varrho\right)~.
% \end{eqnarray}

% A shift of $\sigma$ by $-\imath\Id_{4\gamma/\beta}$ does not change the domain of integration \cite{Guhr2006I,KieburgGroenqvistGuhr} and yields  
% \begin{eqnarray}
%   \label{eq:QandIS}
%   Q(\varrho) &= \exp\left(-\Str \varrho \right)I_{2}(\varrho),
% \end{eqnarray}
% where
% \begin{eqnarray}
% I_{\beta}(\varrho) &= \int \leb{\nu} ~\Sdet^{-\bar n\beta/2}\nu~\exp\left(-\imath \Str \mu \nu\right)  \label{Ingham-Siegel} 
% \end{eqnarray}
% is the supersymmetric Ingham-Siegel Integral. It is a distribution on the space of Hermitian supermatrices, invariant under the adjoint action of $\text{U}(k_1|k_2)$ and its functional form known in the literature \cite{Guhr2006I,KieburgGroenqvistGuhr}.
% 

%The situation becomes even more special if we look at the case discussed in section 
In section \ref{natrualN} it happens that $k_2=\gamma$, $k_1=0$ and $\kappa_{i2}=-t$, $i=1,\dots,\gamma$, \textit{i.e.} Eq.~(\ref{finalZSUSY}) and Eq.~(\ref{Ingham-Siegel}) collapse to integrals over the Fermion-Fermion blocks $\varrho_{FF}$ and $\sigma_{FF}$, respectively, yielding Eq. (\ref{finalESUSY}) and (\ref{backtransIngSieg}).

A more involved situation is that for $\beta=1$ discussed in section \ref{NoNaturalPower}. where the exponent of the determinant is half-integer. We rewrite the integrand as
\begin{eqnarray}
\fl\quad\quad P(W|\Lambda) \Det^{(2\alpha+1)/2}\left(WW\dagg + t\Id_p\right)  &= P(W|\Lambda) \frac{\Det^{\alpha}\left(WW\dagg + t\Id_p\right)}{\Det^{1/2}\left(WW\dagg + t\Id_p\right)}~,
\end{eqnarray}
such that we are in the situation of Eq.~(\ref{definitionZ}) for $\beta=1$. Instead of complex vector, we integrate over a real vector to express the determinant in the denominator as Gaussian integral. The remaining procedure is similar to the general situation and yields the full supermatrix model~(\ref{finalhalfpowerexponent}).
\section*{References}
\bibliography{ref}{}
\bibliographystyle{vancouver}

\end{document}